\documentclass[a4paper,10pt]{article}

\usepackage{amsfonts}
\usepackage{verbatim}
\usepackage{amsmath}
\usepackage{amssymb}
\usepackage{graphicx}
\usepackage{hyperref}

\title{Superembedding methods for 4d $\mathcal{N}$-extended SCFTs}
\author{M. Maio \\~\\ Instituto de F\'\i sica Fundamental, IFF-CSIC, \\~\\ Serrano 123, 28006, Madrid, Spain}

\numberwithin{equation}{section}

\begin{document}

\maketitle

\begin{abstract}
We consider the embedding method of the superconformal group in four dimensions in the case of extended supersymmetry,
hence generalizing the recent work of Goldberger, Skiba and Son which was restricted at $\mathcal{N}=1$. 
Moreover, we work out explicitly the case of $\mathcal{N}=2$ chiral superfields in four dimensions,
putting the component fields in correspondence with Pascal's pyramid at layer $\mathcal{N}$. 
This correspondence is a generic property of the $\mathcal{N}$-extended chiral sector. 
\end{abstract}

\vskip -10cm
\hbox{ \hskip 12cm IFF/FM-2012/05  \hfil}

\clearpage
\tableofcontents

\section{Introduction}
Four-dimensional conformal field theories have attracted much attention in the last decade, 
mainly because of their relevance in the context of the AdS/CFT correspondence and its developments as well as its applications
in completely different fields of which condensed matter is an example.
The conformal symmetry imposes stringent constraints on the theory \cite{Polyakov:1973ha}. 
In the case of two dimensions, for example, 
where scale invariance implies conformal invariance,
the symmetry is enough to solve the theory \cite{Zamolodchikov:1986gt,Belavin:1984vu}. 
In four and higher dimensions, this is in general not true  \cite{Polchinski:1987dy}, 
since examples are known of theories with scale but not conformal symmetry \cite{ElShowk:2011gz,Andreas},
but meaningful statements can be made as well (e.g. \cite{Komargodski:2011vj,Luty:2012ww} for the four-dimensional case).
Moreover, in the presence of supersymmetry, superconformal invariance puts restrictions on the operator scaling dimensions,
independently of the space-time dimensionality \cite{Flato:1983te,Minwalla:1997ka}.

For these theories, a great deal of information can be obtained from the conformal group, that in four
dimensions is $SO(4,2)$ and that acts non-linearly on the coordinates, due to the presence of 
operations such as the inversion and the special conformal transformations.
However, it has been long known that it is possible to formulate the theory in such a way that the conformal
group will act linearly on the coordinates, in the same fashion as the angular momentum does. 
In addition, correlation functions automatically exhibit manifest conformal symmetry
and any results written in terms of the embedding coordinates are valid not just in Minkowski space
but in any conformally flat space-time.
This approach goes back to \cite{Dirac:1936} and had applications to M/string theory branes \cite{Sorokin:1999jx} 
as well as to more complicated conformal field theories in \cite{Mack:1969,Weinberg:2010fx}  
(in particular \cite{Weinberg:2010fx} generalizes the construction from scalars to more generic tensor fields). 
Moreover, embeddings of chiral conformal superspaces were considered in \cite{Siegel:1992-94}, 
in terms of off-shell twistors \cite{Atiyah:1978ri}, 
while \cite{Siegel:2010yd} and \cite{Siegel:2012di} review the general (chiral and non-chiral) case. 
In particular, using the twistor notation, \cite{Siegel:2012di} 
gives the two-point correlators between chiral and anti-chiral superfields for arbitrary $\mathcal{N}$. 
In some cases, $n$-point correlators have been calculated for various values of $n$ 
(see for example \cite{Howe-West} where $\mathcal{N}=4$ and the primaries are not chiral superfields). 
The supertwistor approach was also used in \cite{Kuzenko:2006mv} to study 
the superconformal structure of 4d $\mathcal{N} = 2$ compactified harmonic/projective superspace. 
The results presented here (as well as in \cite{Goldberger:2011yp}) are equivalent to those obtained with 
twistor calculations\footnote{We thank Warren Siegel for this remark.}.

In the embedding method, one introduces two additional coordinates and embeds the four-dimensional space-time into a larger
six-dimensional space with signature $(-,-,+,+,+,+)$. In this space the conformal group generators act as angular momenta.
The four-dimensional space is recovered by constraining the six-dimensional coordinates on the projective light-cone.
The group $SO(4,2)$ is isomorphic to $SU(2,2)$. As a consequence, one can re-express everything in spinor indices 
and introduce gamma matrices for the basis change from $SO(4,2)$ to $SU(2,2)$.

In \cite{Goldberger:2011yp}, the authors generalize the embedding methods to $\mathcal{N}=1$ superconformal field theories.
The superconformal group is $SU(2,2|1)$ and the six-dimensional complex superspace is constructed out of the coordinates 
$(X_{AB},\bar{X}^{AB})$, which contain the standard space-time $X_{\alpha\beta}$, 
one fermionic direction $\theta_\alpha$ and an additional bosonic coordinate $\varphi$. 
$X_{AB}$ transforms linearly under a superconformal transformation.
Upon projecting on the projective light-cone, the four-dimensional superspace $(x^\mu,\theta,\bar{\theta})$ is recovered.
Moreover, scalar and holomorphic fields are considered and it is shown that they correspond to 
four-dimensional $\mathcal{N}=1$ chiral superfields whose $\theta=\bar{\theta}=0$ component is an 
$\mathcal{N}=1$ chiral primary operator \cite{Goldberger:2011yp,Ferrara:1974qk}.

In this paper we generalize the embedding method to extended supersymmetry where the superconformal group is $SU(2,2|\mathcal{N})$.
Most of the arguments are similar to the ones in \cite{Goldberger:2011yp} and analogue results are obtained 
by considering more fermionic coordinates $\theta_{I\alpha}$, with $I=1,\dots,\mathcal{N}$. 
Automatically, more bosonic variables $\varphi_{IJ}$ also appear in the six-dimensional superspace,
which will then be removed once the light-cone constraint is imposed.
We define such a set-up in Section \ref{section: conformal and superconformal group}, where the generators 
as well as the necessary representations are introduced. 
In Section \ref{section: superspace} we construct the $\mathcal{N}$-extended four-dimensional superspace. 
This is done as a coset construction, achieved by applying translations and supersymmetry transformations 
to a reference origin. The resulting space is invariant under the superconformal and Lorentz groups.
In Section \ref{section: the chiral sector} we consider representations of the 
superconformal algebra in terms of superfields. In particular, we generalize the arguments of \cite{Goldberger:2011yp}
to $\mathcal{N}$-extended chiral superfields, find their transformation rules under the $U(\mathcal{N})_R$ symmetry 
subgroup of the superconformal algebra as well as under generic superconformal transformations.
We also study in detail the case of chiral superfields in $\mathcal{N}=2$ extended supersymmetry.

As an aside result of our calculations, we have discovered an interesting connection 
between the number of component fields of a given kind in an $\mathcal{N}$-extended supermultiplet 
and the so-called Pascal's pyramid at layer $\mathcal{N}$. 
This aspect of number theory seems not to be mentioned explicitly in the literature.
One can use it as a check that superfield expansions are correct.

Throughout this paper we use the conventions of \cite{Goldberger:2011yp} and \cite{WB}. In particular, 
the various definitions for the spinor indices and the gamma matrices are the same as in \cite{Goldberger:2011yp}. 
In the appendix we give some details of the calculations as well as our notation. In particular, 
in Appendix \ref{explicit SU(2,2) transformations} we compute explicitly the $SU(2,2)$ transformations 
of the coordinates $X_{\alpha\beta}$.
In Appendix \ref{Appendix useful identities} we give our notation for the spinorial indices and show some 
identities that will be useful in our computations.
In Appendix \ref{4d superspace barred} we provide the necessary for computing the coordinates of the 4d barred superspace.
In Appendix \ref{Appendix Pascal pyramid} we define the Pascal pyramid, with some of its properties and symmetries, 
construct the $\mathcal{N}=4$ chiral multiplet and relate it to the pyramid at layer $4$.

\section{Conformal and superconformal group}
\label{section: conformal and superconformal group}
The superconformal group in four dimensions has fifteen generators that satisfy the $SO(4,2)$ algebra. 
It includes the Poincar\'e generators ($P^\mu$ for translations and $M^{\mu\nu}$ for Lorentz transformations) as a subalgebra.
In addition there are the generators $K^\mu$ for special conformal transformations as well as the dilatation generator $D$.
Special conformal transformations can be thought of as an inversion, followed by a translation and then by another inversion
and act non-linearly on the coordinates $x^\mu$. Schematically:
\begin{subequations}
\begin{eqnarray}
P^\mu & x^\mu\rightarrow x^\mu+a^\mu & \delta x^\mu=a^\mu\\
M^{\mu\nu} & x^\mu\rightarrow \Lambda^\mu_{\phantom{\mu}\nu} x^\nu \quad (\Lambda\in SO(3,1)) & \delta x^\mu=
\omega^\mu_{\phantom{\mu}\nu} x^\nu \qquad (\omega_{\mu\nu}=-\omega_{\nu\mu})\\
K^\mu & x^\mu\rightarrow \frac{x^\mu+x^2b^\mu}{1+2b\cdot x+b^2x^2} & \delta x^\mu=x^2b^\mu-2(b\cdot x)x^\mu\\
D     & x^\mu\rightarrow \lambda x^\mu & \delta x^\mu=\epsilon x^\mu
\end{eqnarray}
\end{subequations}
with real parameters $a^\mu$, $b^\mu$, $\Lambda^\mu_{\phantom{\mu}\nu}=\delta^\mu_\nu+\omega^\mu_{\phantom{\mu}\nu}$ and $\lambda=1+\epsilon$. 
Here indices are raised and lowered by the four-dimensional metric $\eta_{\mu\nu}={\rm diag}(-1,+1,+1,+1)$.

The conformal group $SO(4,2)$ is identical to the Lorentz group in a space with six dimensions and signature $(-,-,+,+,+,+)$.  
One can use this observation to make the conformal transformations act linearly on the coordinates.
Define new variables $X^m$, with $m=+,\mu,-$. Under $SO(4,2)$, $X^m$ transforms linearly:
\begin{equation}
X^m\rightarrow \Lambda^m_{\phantom{m}n}X^n \,,\qquad {\rm with}\quad \Lambda^T\eta\Lambda=\eta \qquad {\rm and} \quad
\eta_{mn}=\left(
\begin{array}{ccc}
0 & 0 & 1/2\\
0 & \eta_{\mu\nu} & 0\\
1/2 & 0 & 0
\end{array}
\right)
\end{equation}
Infinitesimally, $\Lambda^m_{\phantom{m}n}=\delta^m_n+\omega^m_{\phantom{m}n}$, the conformal transformation 
$\delta X^m=\omega^{mn}X_n=\frac{i}{2} \omega^{pq}L_{pq} X^m$ is generated by the $SO(4,2)$ differential operators
\begin{equation}
L^{mn}=iX^m\frac{\partial}{\partial X_n}-iX^n\frac{\partial}{\partial X_m}\,.
\end{equation}
In order to recover the four-dimensional space, one has to demand that the coordinates $X^m$ are constrained on the projective light-cone:
\begin{equation}
X^2\equiv X^m\eta_{mn}X^n=0\,, \qquad X^m\sim\lambda X^m\quad (\lambda\in\mathbb{R})\,.
\end{equation}

Since the conformal group $SO(4,2)$ is equivalent to $SU(2,2)$, we can transform everything into spinor notation. 
$SU(2,2)$ is the group of four by four special matrices that are unitary with respect to a fixed invariant matrix
of signature $\rm diag(+,+,-,-)$.
The connection is realized by using gamma matrices 
to transform the vector index $m$ into an anti-symmetric pair of spinor indices ($\alpha\,,\beta$).
A spinor $V_\alpha$ of $SU(2,2)$ transforms as
\begin{equation}
\label{SU(2,2) spinor transformation}
V_\alpha \rightarrow U_\alpha^{\phantom{\alpha}\beta} V_\beta\,, 
\quad U\in SU(2,2)\,.
\end{equation}
The $SU(2,2)$ index $\alpha=1,\dots,4$ has an undotted component transforming in the fundamental representation of $SL(2,\mathbb{C})$ 
and a dotted component transforming in the complex conjugate representation, $\alpha=(a,\dot{a})$, with $a=1,2$ and $\dot{a}=1,2$ 
(e.g. \cite{WB,Gates:1983nr}). 
The vector $X^m$ becomes an anti-symmetric tensor $X_{\alpha\beta}=-X_{\beta\alpha}$, 
with each index transforming as in (\ref{SU(2,2) spinor transformation}):
\begin{equation}
X^m=\frac{1}{2}X_{\alpha\beta}\Gamma^{m\alpha\beta}\,,
\qquad
X_{\alpha\beta}=\frac{1}{2}X_m\tilde{\Gamma}^m_{\alpha\beta}\,,
\end{equation}
with
\begin{equation}
\tilde{\Gamma}^m_{\alpha\beta}=\frac{1}{2}\epsilon_{\alpha\beta\gamma\delta}(\Gamma^m)^{\gamma\delta}\,.
\end{equation}
The light-cone constraint is written as
\begin{equation}
X_{\alpha\beta}X^{\alpha\beta}=0\,
\end{equation}
with $X^{\alpha\beta}=\frac{1}{2}\epsilon^{\alpha\beta\gamma\delta} X_{\gamma\delta}$. 
Similarly, one can express the $SO(4,2)$ generators $L^{mn}$ in spinor notation. The result is 
$L_\alpha^{\phantom{\alpha}\beta}=-\frac{1}{2}(\Sigma^{mn})_{\alpha}^{\phantom{\alpha}\beta}L_{mn}$, 
$L^{mn}=-(\Sigma^{mn})_{\alpha}^{\phantom{\alpha}\beta}L_\beta^{\phantom{\beta}\alpha}$, 
with $(\Sigma^{mn})_{\alpha}^{\phantom{\alpha}\beta}=-\frac{i}{4}\left(
\tilde{\Gamma}^m\Gamma^n-\tilde{\Gamma}^n\Gamma^n\right)_{\alpha}^{\phantom{\alpha}\beta}$
being the $SU(2,2)$ generators.

\subsection{Superconformal algebra}
The $\mathcal{N}$-extended superconformal group in four dimensions is $SU(2,2|\mathcal{N})$ consisting of matrices of the form
\begin{equation}
\mathcal{U}_A^{\phantom{A}B}=\left(
\begin{array}{cc}
U_\alpha^{\phantom{\alpha}\beta} & \varphi_\alpha^{J}\\
\varkappa_I^{\beta} & z_I^{\phantom{I}J}
\end{array}
\right)\,.
\end{equation}
This is a block matrix, with the indices running in the range $\alpha,\,\beta=1,\dots,4$ and $I,\,J=1,\dots,\mathcal{N}$. 
The diagonal blocks are commuting, while the off-diagonal ones are anti-commuting.
An $SU(2,2|\mathcal{N})$ vector in the fundamental representation is written as
\begin{equation}
V_A=\left(
\begin{array}{c}
V_\alpha\\
\psi_I
\end{array}
\right)\,,
\end{equation}
where the $V_\alpha$ component is fermionic and the $\mathcal{N}$ $\psi_I$ fields are bosonic.
Indices are raised and lowered with the invariant matrix
\begin{equation}
\label{invariant matrix}
A^{\dot{A}B}=\left(
\begin{array}{cc}
A^{\dot{\alpha}\beta} & 0\\
0 & \delta_I^{\phantom{I}J}
\end{array}
\right)\,,
\end{equation}
where $A^{\dot{\alpha}\beta}$ is the $SU(2,2)$ invariant metric.
In order to be an element of $SU(2,2|\mathcal{N})$, the matrix $\mathcal{U}_A^{\phantom{A}B}$ must have unit superdeterminant: 
\begin{equation}
(S\det U)^{-1}=(\det z)^{-1}\cdot 
\det(U-\varphi\cdot z^{-1}\cdot\varkappa)=1\,.
\end{equation}
It must also satisfy
\begin{equation}
\label{definition of SU(2,2|N)}
\mathcal{U}_{\dot{B}}^{\phantom{\dot{B}}\dot{A}} A^{\dot{B}B} \mathcal{U}_B^{\phantom{B}A} = A^{\dot{A}A}\,,
\end{equation}
where $\mathcal{U}_{\dot{B}}^{\phantom{\dot{B}}\dot{A}}=(\mathcal{U}_A^{\phantom{A}B})^\dagger$.

One can introduce the $SU(2,2|\mathcal{N})$ generators $\mathcal{T}_A^{\phantom{A}B}$ as
\begin{equation}
\mathcal{U}_A^{\phantom{A}B}=\delta_A^{\phantom{A}B}+i \mathcal{T}_A^{\phantom{A}B}\,.
\end{equation}
Hence:
\begin{equation}
\mathcal{T}_A^{\phantom{A}B}=\left(
\begin{array}{cc}
T_\alpha^{\phantom{\alpha}\beta} & \phi_\alpha^{J}\\
\psi_I^{\beta} & \phi_I^{\phantom{I}J}
\end{array}
\right)\,,
\end{equation}
with $U_\alpha^{\phantom{\alpha}\beta}=\delta_\alpha^{\phantom{\alpha}\beta}+iT_\alpha^{\phantom{\alpha}\beta}$, 
$z_I^{\phantom{I}J}=\delta_I^{\phantom{I}J}+i\phi_I^{\phantom{I}J}$, 
$\phi_\alpha^{J}=-i\varphi_\alpha^{J}$ and
$\psi_I^{\beta}=-i\varkappa_I^{\beta}$.
From the superdeterminant, it follows that $\mathcal{T}_A^{\phantom{A}B}$ have vanishing supertrace:
\begin{equation}
Str \mathcal{T}= T_\alpha^{\phantom{\alpha}\alpha}-\phi=0\,,
\end{equation}
where $\phi$ is the trace of $\phi_I^{\phantom{I}J}$, $\phi\equiv\sum_I\phi_I^{\phantom{I}I}$.
From (\ref{definition of SU(2,2|N)}), it follows that
\begin{equation}
A^{\dot{A}B} \mathcal{T}_B^{\phantom{B}A} - \mathcal{T}_{\dot{B}}^{\phantom{\dot{B}}\dot{A}} A^{\dot{B}A}=0\,.
\end{equation}
Hence, explicitly
\begin{itemize}
\item for $(A,B)=(\alpha,\beta)$: $A^{\dot{\alpha}\beta} T_{\beta}^{\phantom{\beta}\alpha} - 
T_{\dot{\beta}}^{\phantom{\dot{\beta}}\dot{\alpha}} A^{\dot{\beta}\alpha}=0$,
which implies that $T_\alpha^{\phantom{\alpha}\beta}$ is a $U(2,2)$ generator
\item for mixed indices: $\psi_I^{\alpha}=A^{\dot{\beta}\alpha} \phi_{\dot{\beta}}^{I} 
\equiv \bar{\phi}_I^{\alpha}$ (equivalently $\psi_I=\bar{\phi}_I$)
\item for $(A,B)=(I,J)$: $\phi_I^{\phantom{I}J}=(\phi^\dagger)_I^{\phantom{I}J}$ (equivalently $\phi=\phi^\dagger$), 
which states that $\phi_I^{\phantom{I}J}$ is hermitian and hence $z_I^{\phantom{I}J}=(e^{i\phi})_I^{\phantom{I}J}$ is unitary, 
$z_I^{\phantom{I}J}\in U(\mathcal{N})$, as it should be since $\phi_I^{\phantom{I}J}$ is the $R$-symmetry generator. 
\end{itemize}
In order to make $T_\alpha^{\phantom{\alpha}\beta}$ an $SU(2,2)$ generator 
(and hence $U_\alpha^{\phantom{\alpha}\beta}\in SU(2,2)$), one can subtract its trace and replace $\mathcal{T}_A^{\phantom{A}B}$ by
\begin{equation}
\mathcal{T}_A^{\phantom{A}B}=\left(
\begin{array}{cc}
T_\alpha^{\phantom{\alpha}\beta}+\frac{1}{4}\delta_\alpha^{\phantom{\alpha}\beta}\phi & \phi_\alpha^{J}\\
\bar{\phi}_I^{\beta} & \phi_I^{\phantom{I}J}
\end{array}
\right)
\end{equation}
in the fundamental representation.

By defining supercharges according to 
\begin{equation}
-i\delta V_A= [\bar{\phi}_J^{\beta} Q_\beta^J + \phi_{\alpha}^{J} \bar{Q}_J^\beta] V_A\,,
\end{equation}
one has
\begin{equation}
(Q_\sigma^L)_A^{\phantom{A}B}=\left(
\begin{array}{cc}
0 & 0\\
\delta_\sigma^{\phantom{\sigma}\beta} \delta_I^{\phantom{I}L} & 0
\end{array}
\right)
\qquad {\rm and} \qquad
(\bar{Q}^\sigma_L)_A^{\phantom{A}B}=\left(
\begin{array}{cc}
0 & \delta_\alpha^{\phantom{\alpha}\sigma} \delta_L^{\phantom{L}J}\\
0 & 0
\end{array}
\right)\,.
\end{equation}
Similarly, the $R$-symmetry generator is a non-abelian matrix
\begin{equation}
R_A^{\phantom{A}B}=\left(
\begin{array}{cc}
\frac{1}{4}\delta_\alpha^{\phantom{\alpha}\beta} \phi & 0\\
0 & \phi_I^{\phantom{I}J}
\end{array}
\right)\,,
\end{equation}
which contains the $U(\mathcal{N})$ block of the extended supersymmetry.

\subsection{Representations and invariants}
Many concepts here are similar to those presented in \cite{Goldberger:2011yp}. If $V_A$ transforms in the fundamental representation, 
\begin{equation}
V_A=\left(
\begin{array}{c}
V_\alpha\\
\psi_I
\end{array}
\right)\,,
\qquad V_A\rightarrow \mathcal{U}_A^{\phantom{A}B} V_B
\end{equation}
and $\bar{W}^A\equiv(W^\dagger)_{\dot{B}}A^{\dot{B}A}$ in the anti-fundamental, 
\begin{equation}
\bar{W}^A=\left(
\begin{array}{c}
\bar{W}^\alpha\\
\bar{\psi}^I
\end{array}
\right)\,,
\qquad \bar{W}^A\rightarrow \bar{W}^B (\mathcal{U}^{-1})_B^{\phantom{B}A}\,,
\end{equation}
then the product $Z_A^{\phantom{A}B}=V_A \bar{W}^B$ transforms as a tensor
\begin{equation}
Z_A^{\phantom{A}B}\rightarrow \mathcal{U}_A^{\phantom{A}C} Z_C^{\phantom{C}D} (\mathcal{U}^{-1})_D^{\phantom{D}B}\,.
\end{equation}
The singlet representation is given by its supertrace,
\begin{equation}
Str Z_A^{\phantom{A}B} = - Z_A^{\phantom{A}B} \lambda_B^{\phantom{B}A}\,,
\end{equation} 
where
\begin{equation}
\lambda_A^{\phantom{A}B}=\left(
\begin{array}{cc}
-\delta_\alpha^{\phantom{\alpha}\beta} & 0 \\
0 & \delta_I^{\phantom{I}J}
\end{array}
\right)\,,
\end{equation}
which, due to the property of the supertrace $Str(\mathcal{U}_1 \mathcal{U}_2)=Str(\mathcal{U}_2 \mathcal{U}_1)$, 
is invariant under $SU(2,2|\mathcal{N})$ transformations.
The adjoint representation is given by the part of $Z_A^{\phantom{A}B}$ with vanishing supertrace,
\begin{equation}
Z_A^{\phantom{A}B} + \frac{1}{4-\mathcal{N}} (Z_C^{\phantom{C}D} \lambda_D^{\phantom{D}C}) \delta_A^{\phantom{A}B}\,.
\end{equation}

Tensors $X_{AB}$ and $\bar{X}^{AB}$ transform\footnote{To check these transformations, use the inversion formula for block matrices.
One convenient expression is:
\begin{equation}
(\mathcal{U}^{-1})_A^{\phantom{A}B}=\left(
\begin{array}{cc}
\big(U^{-1}+U^{-1}\phi(z-\psi U^{-1}\phi)^{-1}\psi U^{-1}\big)_\alpha^{\phantom{\alpha}\beta} & 
-\big(U^{-1}\phi(z-\psi U^{-1}\phi)^{-1}\big)_\alpha^{J}\\ 
-\big((z-\psi U^{-1}\phi)^{-1}\psi U^{-1}\big)_I^{\beta} & 
\big((z-\psi U^{-1}\phi)^{-1}\big)_I^{\phantom{I}J} 
\end{array}
\right)\,.\nonumber
\end{equation}
}
as the products $X_{AB}\sim V_A V_B$ and $\bar{X}^{AB}\sim \bar{V}^A \bar{V}^B$:
\begin{eqnarray}
\label{superconformal transformations}
X_{AB}&\rightarrow& \mathcal{U}_A^{\phantom{A}C} X_{CD} \hat{\mathcal{U}}_B^{\phantom{B}D}\,, \quad
\hat{\mathcal{U}}_B^{\phantom{B}D}=\left(
\begin{array}{cc}
\mathcal{U}_\alpha^{\phantom{\alpha}\beta} & \mathcal{U}_\alpha^{J}\\
-\mathcal{U}_I^{\beta} & \mathcal{U}_I^{\phantom{I}J}
\end{array}
\right)\\
\bar{X}^{AB}&\rightarrow& (\hat{\mathcal{U}}^{-1})_C^{\phantom{C}A} \bar{X}^{CD} (\mathcal{U}^{-1})_D^{\phantom{D}B}\,, \quad
(\hat{\mathcal{U}}^{-1})_D^{\phantom{D}B}=\left(
\begin{array}{cc}
(\mathcal{U}^{-1})_\alpha^{\phantom{\alpha}\beta} & (\mathcal{U}^{-1})_\alpha^{J}\\
-(\mathcal{U}^{-1})_I^{\beta} & (\mathcal{U}^{-1})_I^{\phantom{I}J}
\end{array}
\right)\,.
\nonumber
\end{eqnarray}
This implies that the product $X_{AC}\lambda_D^{\phantom{D}C}\bar{X}^{DB}$ transforms as $Z_A^{\phantom{A}B}$ and hence the scalar product
\begin{equation}
\label{invariant scalar product}
X\cdot\bar{Y}\equiv Str(X\lambda\bar{Y})=Tr(-\lambda X\lambda\bar{Y})
\end{equation}
is $SU(2,2|\mathcal{N})$ invariant.

\section{Superspace}
\label{section: superspace}
The tensor $X_{AB}$ contains the superspace components of six-dimensional conformal group. Explicitly:
\begin{equation}
X_{AB}=\left(
\begin{array}{cc}
X_{\alpha\beta} & \theta_{J\alpha}\\
\theta_{I\beta} & \varphi_{IJ}
\end{array}
\right)\,,
\end{equation}
where $X_{\alpha\beta}$ is anti-symmetric, $\theta_{I\alpha}\equiv X_{I\alpha}=X_{\alpha I}$ are fermionic variables 
and $\varphi_{IJ}\equiv X_{IJ}$ is a symmetric matrix of bosonic coordinates.
In order to use the invariant scalar product (\ref{invariant scalar product}), we need the conjugate of $X$, $\bar{X}^{AB}$, with components
\begin{equation}
\bar{X}^{AB}=\left(
\begin{array}{cc}
\bar{X}^{\alpha\beta} & \bar{\theta}^{J\alpha}\\
\bar{\theta}^{I\beta} & \bar{\varphi}^{IJ}
\end{array}
\right)\,.
\end{equation}
Defining $X_{\dot{\alpha}\dot{\beta}}=(X_{\alpha\beta})^\dagger$ and $\theta^I_{\dot{\alpha}}=(\theta_{I\alpha})^\dagger$, 
and using the invariant matrix (\ref{invariant matrix}), 
the components are $\bar{X}^{\alpha\beta}=A^{\dot{\alpha}\alpha}A^{\dot{\beta}\beta}X_{\dot{\alpha}\dot{\beta}}$, 
$\bar{\theta}^{I\alpha}=A^{\dot{\alpha}\alpha}\theta^I_{\dot{\alpha}}$, $\bar{\varphi}^{IJ}=(\varphi_{IJ})^\dagger$.
The full superspace is then described by the supercoordinates $(X_{AB},\bar{X}^{AB})$, 
endowed with the scalar product (\ref{invariant scalar product}).
In components:
\begin{equation}
X_1\cdot \bar{X}_2=X_{1\alpha\beta}\bar{X}_2^{\alpha\beta}+2\theta_{1I\alpha}\bar{\theta}_2^{I\alpha}-\varphi_{1JI}\bar{\varphi}_2^{\phantom{2}IJ}\,.
\end{equation}
The superconformal transformations of the coordinates can be computed by using (\ref{superconformal transformations}). 
For $X_{AB}$, they are
$-i\delta X_{AB}=X_{AC}\hat{\mathcal{T}}_B^{\phantom{B}C}+\mathcal{T}_A^{\phantom{A}C}X_{CB}$, or in components:
\begin{subequations}
\label{superconformal transformations for superspace}
\begin{eqnarray}
-i\delta X_{\alpha\beta}&=& T_\alpha^{\phantom{\alpha}\gamma}X_{\gamma\beta}+T_\beta^{\phantom{\beta}\gamma}X_{\alpha\gamma}
+ \phi_\alpha^I\theta_{I\beta}-\phi^I_\beta\theta_{I\alpha} +\frac{1}{2}\phi X_{\alpha\beta}\\
-i\delta \theta_{I\alpha}&=& T_\alpha^{\phantom{\alpha}\gamma}\theta_{I\gamma}+ \bar{\phi}_I^{\beta}X_{\beta\alpha}
+ \phi_\alpha^{J}\varphi_{JI}+\phi_I^{\phantom{I}J}\theta_{J\alpha}+\frac{1}{4}\phi\theta_{I\alpha}\\
-i\delta \varphi_{IJ}&=& \bar{\phi}_J^{\gamma}\theta_{I\gamma}+\bar{\phi}_I^{\gamma}\theta_{J\gamma}
+ \phi_J^{\phantom{J}K}\varphi_{IK}+\phi_I^{\phantom{I}K}\varphi_{KJ}
\end{eqnarray}
\end{subequations}
For $\bar{X}^{AB}$, they are $i\delta \bar{X}^{AB}=\bar{X}^{AC}\mathcal{T}_C^{\phantom{D}B}+\hat{\mathcal{T}}_C^{\phantom{C}A}\bar{X}^{CB}$, 
or in components:
\begin{subequations}
\label{superconformal transformations for superspace bar}
\begin{eqnarray}
i\delta \bar{X}^{\alpha\beta}&=& T^{\phantom{\gamma}\alpha}_\gamma\bar{X}^{\gamma\beta}+T^{\phantom{\gamma}\beta}_\gamma\bar{X}^{\alpha\gamma}
+ \bar{\phi}^\alpha_I\bar{\theta}^{I\beta}-\bar{\phi}_I^\beta\bar{\theta}^{I\alpha} +\frac{1}{2}\phi \bar{X}^{\alpha\beta}\\
i\delta \bar{\theta}^{I\alpha}&=& T_\gamma^{\phantom{\gamma}\alpha}\bar{\theta}^{I\gamma}-\phi_\gamma^{I}\bar{X}^{\gamma\alpha}
+ \bar{\phi}_J^{\alpha}\bar{\varphi}^{IJ}+\phi_J^{\phantom{J}I}\bar{\theta}^{J\alpha}+\frac{1}{4}\phi\bar{\theta}^{I\alpha}\\
i\delta \bar{\varphi}^{IJ}&=& -\phi_\gamma^{I}\bar{\theta}^{J\gamma}-\phi_\gamma^{J}\bar{\theta}^{I\gamma}
+ \phi_K^{\phantom{K}I}\bar{\varphi}^{KJ}+\phi_K^{\phantom{K}J}\bar{\varphi}^{IK}
\end{eqnarray}
\end{subequations}

\subsection{4D superspace}
In this section we construct the four-dimensional superspace, described by the projective coordinates 
$(X_{AB},\bar{X}^{AB})\sim (\lambda X_{AB},\bar{\lambda}\bar{X}^{AB})$, $\lambda\in\mathbb{C}$, on the six-dimensional light-cone
\begin{equation}
X\cdot\bar{X}=0\,.
\end{equation}
We start from the origin defined as
\begin{equation}
\label{origin in spinor notation}
\hat{X}_{\alpha\beta}=\frac{1}{2}\left(
\begin{array}{cc}
i\epsilon_{ab}X^+ & 0\\
0 & 0
\end{array}
\right)\,,\quad
\hat{\theta}_{I\alpha}=0 \,,\quad \hat{\varphi}_{IJ}=0\,.
\end{equation}
One can check that the origin satisfies the light-cone constraint\footnote{Since:
\begin{equation}
\hat{\bar{X}}^{\alpha\beta}=\frac{1}{2}\left(
\begin{array}{cc}
0 & 0\\
0 & i\epsilon_{\dot{a}\dot{b}}\bar{X}^+
\end{array}
\right)\,.
\end{equation}
}.
An arbitrary point in superspace is reached by applying all possible superconformal transformations 
(\ref{superconformal transformations for superspace}) and (\ref{superconformal transformations for superspace bar}).
Under superconformal transformations, the origin transforms as
\begin{subequations}
\label{transf of the origin}
\begin{eqnarray}
-i\delta \hat{X}_{\alpha\beta}&=& \frac{1}{2}i X^+
\left(\delta_\beta^b T_\alpha^{\phantom{\alpha}c}\epsilon_{cb}+ \delta_\alpha^a T_\beta^{\phantom{b}c}\epsilon_{ac}+
\delta_\alpha^a \delta_\beta^b\frac{1}{2}\epsilon_{ab}\phi\right)\\
-i\delta \hat{\theta}_{I\alpha}&=& -\delta_\alpha^a \frac{1}{2}i \epsilon_{ab}\bar{\phi}_I^{b}X^+\\
-i\delta \hat{\varphi}_{IJ}&=& 0\,.
\end{eqnarray}
\end{subequations}
Explicitly, from table \ref{transformations of the origin} in Appendix \ref{explicit SU(2,2) transformations}, 
the origin is invariant under special conformal and Lorentz transformations, and projectively invariant under dilatations. 
It is also invariant under supersymmetry transformations generated by a spinor of the form
\begin{equation}
\phi_\alpha^{I}\sim\left(
\begin{array}{c}
\eta_a^{I}\\ 
0
\end{array}
\right)\,,
\end{equation}
because in this case $\bar{\phi}_I^\beta$ in (\ref{transf of the origin}) has only the down (dotted) component.
Hence a generic point in superspace is reached from the origin by applying translations with parameter $\omega^{\mu-}=2\delta x^\mu=2x^\mu$, 
and supersymmetry transformations with spinorial parameter
\begin{equation}
\phi_\alpha^{I}=\left(
\begin{array}{c}
0\\
2\bar{\theta}^{I\dot{a}}
\end{array}
\right)\,.
\end{equation}
Then the four-dimensional superspace is a coset space where points are labelled by 
$(x^\mu,\{\theta_{Ia}\},\{\bar{\theta}^{J\dot{a}}\})$ and the symmetries above have been modded out.
We denote the whole set of theta coordinates by curly brackets: 
\begin{equation}
\{\theta_{Ia}\}=\{\theta_{1a},\theta_{2a},\dots,\theta_{\mathcal{N}a}\}
\qquad {\rm and} \qquad
\{\bar{\theta}^{I\dot{a}}\}=\{\bar{\theta}^{1\dot{a}},\bar{\theta}^{2\dot{a}},\dots,\bar{\theta}^{\mathcal{N}\dot{a}}\}\,.
\end{equation}
The explicit transformation that brings the origin to the point $(x^\mu,\{\theta_{Ia}\},\{\bar{\theta}^{J\dot{a}}\})$ in superspace
is a product of two commuting transformations, one in ordinary space-time and the other one in the theta directions:
\begin{equation}
\label{full transformation in 4d superspace}
\mathcal{U}(x^\mu,\{\theta_{Ia}\},\{\bar{\theta}^{J\dot{a}}\})_A^{\phantom{A}B}=
\mathcal{U}(x^\mu,0,0)_A^{\phantom{A}C} 
\mathcal{U}(0,\{\theta_{Ia}\},\{\bar{\theta}^{J\dot{a}}\})_C^{\phantom{C}B}\,.
\end{equation}
The space-time part is as in \cite{Goldberger:2011yp}:
\begin{equation}
\label{space-time factor in the product}
\mathcal{U}(x^\mu,0,0)_A^{\phantom{A}B}=
\left(
\begin{array}{cc}
\left(
e^{-i x_\mu \Sigma^{\mu +}}
\right)_\alpha^{\phantom{\alpha}\beta}
& 0\\
0 & \delta_I^{\phantom{I}J}
\end{array}
\right)=
\left(
\begin{array}{ccc}
\delta_a^{\phantom{a}b}&0&0\\
ix^\mu(\bar{\sigma}_\mu)^{\dot{a}b}&\delta^{\dot{a}}_{\phantom{\dot{a}}\dot{b}}&0\\
0&0&\delta_I^{\phantom{I}J}
\end{array}
\right)\,,
\end{equation}
where $\delta_I^{\phantom{I}J}$ is the $\mathcal{N}\times\mathcal{N}$ identity matrix. 
The theta part is a generalization of the expression used in \cite{Goldberger:2011yp} (see Appendix \ref{Appendix useful identities}): 
\begin{equation}
\label{theta factor in the product}
\mathcal{U}(0,\{\theta_{Ia}\},\{\bar{\theta}^{J\dot{a}}\})_A^{\phantom{A}B}=
\left(
\begin{array}{cc}
\delta_\alpha^{\phantom{\alpha}\beta}-\frac{1}{2}\sum_I\phi^I_\alpha\bar{\phi}^\beta_I & i\phi_\alpha^J\\
i\bar{\phi}^\beta_I & \delta_I^{\phantom{I}J}
\end{array}
\right)=
\left(
\begin{array}{ccc}
\delta_a^{\phantom{a}b}&0&0\\
-\sum_I\theta_I\sigma^\mu\bar{\theta}^I(\bar{\sigma}_\mu)^{\dot{a}b}&\delta^{\dot{a}}_{\phantom{\dot{a}}\dot{b}}&2i\bar{\theta}^{J\dot{a}}\\
2i\theta_I^{b}&0&\delta_I^{\phantom{I}J}
\end{array}
\right)\,.
\end{equation}
This matrix is also equal to the product of analogous matrices, but with only one theta coordinate different from zero:
\begin{equation}
\mathcal{U}(0,\{\theta_{Ia}\},\{\bar{\theta}^{J\dot{a}}\})=
\prod_{I=1}^{\mathcal{N}}
\mathcal{U}(0,\theta_{Ia},\bar{\theta}^{I\dot{a}})=
\mathcal{U}(0,\theta_{1a},\bar{\theta}^{1\dot{a}})
\mathcal{U}(0,\theta_{2a},\bar{\theta}^{2\dot{a}})\cdots 
\mathcal{U}(0,\theta_{\mathcal{N}a},\bar{\theta}^{\mathcal{N}\dot{a}})\,,
\end{equation}
where in the $I$-th factor $\theta_{Ja}=\bar{\theta}^{J\dot{a}}=0$ if $J \neq I$\footnote{
E.g. for $\mathcal{N}=2$ one has:
\begin{eqnarray}
&&\left(
\begin{array}{cccc}
\delta_a^{\phantom{a}b}&0&0&0\\
-\sum_I\theta_I\sigma^\mu\bar{\theta}^I(\bar{\sigma}_\mu)^{\dot{a}b}&\delta^{\dot{a}}_{\phantom{\dot{a}}\dot{b}}
&2i\bar{\theta}^{1\dot{a}}&2i\bar{\theta}^{2\dot{a}}\\
2i\theta_1^{b}&0&1&0\\
2i\theta_2^{b}&0&0&1
\end{array}
\right)
=\nonumber
\\&&\qquad\qquad=
\left(
\begin{array}{cccc}
\delta_a^{\phantom{a}b}&0&0&0\\
-\theta_1\sigma^\mu\bar{\theta}^1(\bar{\sigma}_\mu)^{\dot{a}b}&\delta^{\dot{a}}_{\phantom{\dot{a}}\dot{b}}&2i\bar{\theta}^{1\dot{a}}&0\\
2i\theta_1^{b}&0&1&0\\
0&0&0&1
\end{array}
\right)
\cdot
\left(
\begin{array}{cccc}
\delta_a^{\phantom{a}b}&0&0&0\\
-\theta_2\sigma^\mu\bar{\theta}^2(\bar{\sigma}_\mu)^{\dot{a}b}&\delta^{\dot{a}}_{\phantom{\dot{a}}\dot{b}}&0&2i\bar{\theta}^{2\dot{a}}\\
0&0&1&0\\
2i\theta_2^{b}&0&0&1
\end{array}
\right)\,.
\nonumber
\end{eqnarray}
}.
One can explicitly check that each of the single factors commutes with each other as well as
with the space-time transformation (\ref{space-time factor in the product}) and that, consequently, 
the full theta matrix (\ref{theta factor in the product}) commutes with (\ref{space-time factor in the product}) too.
The complete transformation from the origin to any point in the superspace is explicitly
\begin{equation}
\label{explicit full transformation in 4d superspace}
\mathcal{U}(x^\mu,\theta_{Ia},\bar{\theta}^{J\dot{a}})_A^{\phantom{A}B}=\left(
\begin{array}{ccc}
\delta_a^{\phantom{a}b}&0&0\\
iy^\mu(\bar{\sigma}_\mu)^{\dot{a}b}&\delta^{\dot{a}}_{\phantom{\dot{a}}\dot{b}}&2i\bar{\theta}^{J\dot{a}}\\
2i\theta_I^{b}&0&\delta_I^{\phantom{I}J}
\end{array}
\right)\,,
\end{equation}
where
\begin{equation}
\label{chiral coordinate}
y^\mu=x^\mu+i\sum_I\theta_I\sigma^\mu\bar{\theta}^I 
\end{equation}
is the coordinate of the $\mathcal{N}$-extended chiral superspace\footnote{See Section 3 (formula (3.3.28) in particular) 
of \cite{Gates:1983nr} for any $\mathcal{N}$, or e.g. \cite{Lykken:1996xt} explicitly for $\mathcal{N}=2$.}.

We use eq. (\ref{explicit full transformation in 4d superspace}) to deduce the coordinates of the 4d superspace. 
Starting from the origin (\ref{origin in spinor notation}) and using (\ref{superconformal transformations}), 
or equivalently (\ref{superconformal transformations for superspace})-(\ref{superconformal transformations for superspace bar}), 
one has\footnote{See Appendix \ref{Appendix useful identities}.}:
\begin{subequations}
\label{Coordinates of 4d superspace}
\begin{eqnarray}
X_{\alpha\beta}(y,\theta)&=&\mathcal{U}_\alpha^{\phantom{\alpha}\gamma}\hat{X}_{\gamma\delta}\mathcal{U}_\beta^{\phantom{\beta}\delta}=
\frac{i}{2}X^+\left(
\begin{array}{cc}
\epsilon_{ab} & \epsilon_{ad}iy^\mu(\bar{\sigma}_\mu)^{\dot{b}d}\\
iy^\mu(\bar{\sigma}_\mu)^{\dot{a}c}\epsilon_{cb} & -y^\mu y^\nu(\bar{\sigma}_\mu)^{\dot{a}c}(\bar{\sigma}_\nu)^{\dot{b}d}\epsilon_{cd}
\end{array}
\right)
\equiv\frac{1}{2}Y_m\tilde{\Gamma}^m_{\alpha\beta}\\
\theta_{I\alpha}(y,\theta)&=&-\mathcal{U}_\alpha^{\phantom{\alpha}\gamma}\hat{X}_{\gamma\delta}\mathcal{U}_I^{\phantom{\beta}\delta}=X^+\left(
\begin{array}{c}
\theta_{Ia}\\
iy^\mu(\bar{\sigma}_\mu)^{\dot{a}b}\theta_{Ib}
\end{array}
\right)\\
\varphi_{IJ}(y,\theta)&=&-\mathcal{U}_I^{\phantom{\alpha}\gamma}\hat{X}_{\gamma\delta}\mathcal{U}_J^{\phantom{\beta}\delta}=
2iX^+ \theta_I \cdot \theta_J\,.
\end{eqnarray}
\end{subequations}
where 
\begin{equation}
Y^m=(Y^+=X^+, Y^\mu=X^+ y^\mu, Y^-=-X^+ y^2)
\end{equation}
and we use the epsilon tensor to lower spinorial indices, 
$\theta_{Ia}=\epsilon_{ab}\theta_I^{b}$, so that $\theta_I \cdot \theta_J=\theta_I^{c} \epsilon_{cd} \theta_J^{d}=\theta_J \cdot \theta_I$. 

Similarly, for the barred coordinates one finds (more details in Appendix \ref{4d superspace barred}):
\begin{subequations}
\label{barred 4d superspace}
\begin{eqnarray}
\bar{X}^{\alpha\beta}&=&(\mathcal{U}^{-1})_{\dot{c}}^{\phantom{\dot{c}}\alpha}\hat{\bar{X}}^{\dot{c}\dot{d}}
(\mathcal{U}^{-1})_{\dot{d}}^{\phantom{\dot{d}}\beta}
=\frac{1}{2}\bar{Y}_m\Gamma^{m\alpha\beta}\\
\bar{\theta}^{I\alpha}&=&(\mathcal{U}^{-1})_{\dot{c}}^{\phantom{\dot{c}}I}\hat{\bar{X}}^{\dot{c}\dot{d}}
(\mathcal{U}^{-1})_{\dot{d}}^{\phantom{\dot{d}}\alpha}
=\bar{X}^+\left(
\begin{array}{c}
-i\bar{y}^\mu(\bar{\sigma}_\mu)^{\dot{b}a}\bar{\theta}^I_{\dot{b}}\\
\bar{\theta}^I_{\dot{a}}
\end{array}
\right)\\
\bar{\varphi}^{IJ}&=&(\mathcal{U}^{-1})_{\dot{c}}^{\phantom{\dot{c}}I}\hat{\bar{X}}^{\dot{c}\dot{d}}
(\mathcal{U}^{-1})_{\dot{d}}^{\phantom{\dot{d}}\alpha}
=-2i\bar{X}^+ \bar{\theta}^I \cdot \bar{\theta}^J\,,
\end{eqnarray}
\end{subequations}
where 
\begin{equation}
\bar{Y}^m=(\bar{Y}^+=\bar{X}^+, \bar{Y}^\mu=\bar{X}^+ \bar{y}^\mu, Y^-=-\bar{X}^+ \bar{y}^2)=(Y^m)^\dagger\,,
\end{equation}
$\bar{y}^\mu=x^\mu-i\sum_I \theta_I\sigma^\mu\bar{\theta}^I$ and we have lowered the theta index, 
$\bar{\theta}^I_{\dot{a}}=\epsilon_{\dot{a}\dot{b}}\bar{\theta}^{I\dot{b}}$.

Under superconformal transformations, $(y^\mu,\theta_a)$ transform as the coordinates of the $\mathcal{N}$-extended chiral superspace. 
In fact, using (\ref{superconformal transformations for superspace}) with
\begin{equation}
\label{superconformal parameter}
\phi^I_\alpha=\left(
\begin{array}{c}
-2\eta^I_a\\
2\bar{\epsilon}^{I\dot{a}}
\end{array}
\right)
\Longleftrightarrow
\bar{\phi}_I^\alpha=(2\epsilon^a_I,-2\bar{\eta}_{I\dot{a}})
\end{equation}
as the only non-vanishing parameter, we get the following transformations:
\begin{subequations}
\label{4d space-time superconformal transformations}
\begin{eqnarray}
\delta y^\mu&=&\sum_I(2i\theta_I\sigma^\mu\bar{\epsilon}^I-2y_\nu\theta_I\sigma^\mu\bar{\sigma}^\nu\eta^I)\,,\\
\delta \theta_I^{a}&=&\epsilon_I^{a}-iy^\mu(\bar{\sigma}_\mu)^{\dot{b}a}\bar{\eta}^I_{\dot{b}}+4\sum_J\theta_I\cdot\theta_J\eta^{Ja}\,.
\end{eqnarray}
\end{subequations}
Including also the barred half superspace given by $\bar{X}^{AB}$, one gets the full superspace
spanned by the coordinates $(x=(y+\bar{y})/2,\theta_I,\bar{\theta}_I)$.

\section{The chiral sector}
\label{section: the chiral sector}
$\mathcal{N}=1$ conformal chiral superfields, $\Phi(y^\mu,\theta_a)$, were considered in \cite{Goldberger:2011yp}. 
Defining a chiral superfield on the light-cone
\begin{equation}
\Phi(y^\mu,\theta_a)\equiv(X^+)^\Delta\Phi(X_{AB})\,,
\end{equation}
and expanding in powers of theta
\begin{equation}
\Phi(y^\mu,\theta_a)=A(y)+\sqrt{2}\theta\psi(y)+\theta^2F(y)\,,
\end{equation}
with $y^\mu=x^\mu+i\theta\sigma^\mu\bar{\theta}$,
it was checked that the component fields transform as a chiral multiplet under the $\mathcal{N}=1$ super Poincar\'e group
and have non-trivial rules under the special conformal transformations. In particular,
the highest component $A(0)$ is a chiral primary operator, in the sense that it is annihilated by the special superconformal 
generators $S_a$ and $\bar{S}^{\dot{a}}$, which are related to the supercharges by
\begin{equation}
Q_\alpha=\frac{i}{2}\left(
\begin{array}{c}
-Q_a \\ \bar{S}^{\dot{a}}
\end{array}
\right)
\qquad {\rm and} \qquad
\bar{Q}^\alpha=-\frac{i}{2}\left(
\begin{array}{c}
S^a \\ -\bar{Q}_{\dot{a}}
\end{array}
\right)\,.
\end{equation}
Moreover, under the $U(1)_R\subset SU(2,2|1)$ $R$-symmetry, one recovers the fact 
that the $R$-charge is proportional to the scaling dimension $\Delta$ at a superconformal fixed point \cite{Goldberger:2011yp}, 
according to 
\begin{equation}
\Phi'(y,e^{3i\phi/4}\theta)=e^{i\frac{\phi}{2}\Delta}\Phi(y,\theta)\,.
\end{equation}
Actually, this result continues to hold in the case of extended supersymmetry. Under $U(\mathcal{N})_R$ the indices $I,J,\dots$
transform with a unitary matrix:
\begin{subequations}
\begin{eqnarray}
-i\delta X_{\alpha\beta}=\frac{1}{2}\phi X_{\alpha\beta}&\rightarrow&
X'_{\alpha\beta}=e^{i\frac{\phi}{2}}X_{\alpha\beta}\\ 
-i\delta \theta_{I\alpha}=\phi_I^{\phantom{I}J}\theta_{J\alpha}+\frac{1}{4}\phi\theta_{I\alpha}&\rightarrow&
\theta'_{I\alpha}=e^{i\frac{\phi}{4}}\left(e^{i\phi}\right)_I^{\phantom{I}J}\theta_{J\alpha}\\
-i\delta \varphi_{IJ}=\phi_J^{\phantom{J}K}\varphi_{IK}+\phi_I^{\phantom{I}K}\varphi_{KJ}&\rightarrow&
\varphi'_{IJ}=\left(e^{i\phi}\right)_I^{\phantom{I}K}\left(e^{i\phi}\right)_J^{\phantom{J}L}\varphi_{KL}\,.
\end{eqnarray}
\end{subequations}
On the light-cone we define 
\begin{equation}
\Phi(y,\theta_{Ia})\equiv(X^+)^\Delta \Phi(X_{\alpha\beta},\theta_{I\alpha},\varphi_{IJ})\,,
\end{equation}
where the components are functions of ($y,\theta_{Ia}$). 
Hence, as $\Phi$ is a scalar, i.e. 
\begin{equation}
\Phi'(X'_{\alpha\beta},\theta'_{I\alpha},\varphi'_{IJ})=\Phi(X_{\alpha\beta},\theta_{I\alpha},\varphi_{IJ})\,,
\end{equation}
we find
\begin{equation}
\Phi'(y',\theta'_{Ia})=
\left(e^{i\frac{\phi}{2}}X^+\right)^\Delta\Phi'(X'_{\alpha\beta},\theta'_{I\alpha},\varphi'_{IJ})=
e^{i\frac{\phi}{2}\Delta}\Phi(y,\theta_{Ia})\,,
\end{equation}
under $U(\mathcal{N})_R\subset SU(2,2|\mathcal{N})$.
This result is the same as in the $\mathcal{N}=1$ case, as we wanted to show. In the above equation we have used that 
$X'^+=e^{i\frac{\phi}{2}}X^+$ and hence $y'=y$, $\theta'_{Ia}=e^{i\frac{3}{4}\phi}\left(e^{i\phi}\right)_I^{\phantom{I}J}\theta_{Ja}$,
as well as the projective property $\Phi(\lambda X_{AB})=\lambda^{-\Delta}\Phi(X_{AB})$ with $\lambda=(X^+)^{-1}$.

Under a generic superconformal transformation generated by the spinor (\ref{superconformal parameter}),
using the scaling property of the superfield $\Phi(y,\theta_I)$ as well as its scalar nature, we have
\begin{equation}
\label{component transformations from superfield}
\delta\Phi(y,\theta_I)=\Delta\frac{\delta X^+}{X^+}\Phi(y,\theta_I)-\delta y^\mu\frac{\partial\Phi}{\partial y^\mu}(y,\theta_I)
-\delta\theta_{I}^a\frac{\partial\Phi}{\partial\theta_{I}^a}(y,\theta_I)
\end{equation}
where $\delta X^+=-4X^+\theta_I\cdot\eta^I$ and the variations
$\delta y^\mu$ and $\delta\theta_{I}^a$ are given by (\ref{4d space-time superconformal transformations}). 
The transformations for the component fields can be read off directly from this one by equating same powers of theta on both sides.
Special superconformal transformations are obtained by setting $\epsilon_{I}^a=0$ in these equations.

$\mathcal{N}$-extended superfields can be represented in terms of $\mathcal{N}=1$ superfields by 
expanding all the theta coordinates but one. In the case of $\mathcal{N}$-extended \emph{chiral} superfields,
the component fields are $\mathcal{N}=1$ \emph{chiral} superfields and one can repeat the above argument.
The four-dimensional expansion strongly depends on the value of $\mathcal{N}$, 
since the number of components grows exponentially with $\mathcal{N}$. 
In the following Subsection we will consider $\mathcal{N}=2$.

\subsection{ \texorpdfstring{$\mathcal{N}=2$}{} chiral superfields}
As an example, consider an $\mathcal{N}=2$ chiral superfield, that we denote by $\Psi(x,\theta_1,\bar{\theta}^1,\theta_2,\bar{\theta}^2)$.
The solution to the the chiral constraints $\bar{D}_{1\dot{a}}\Psi=0$ and $\bar{D}_{2\dot{a}}\Psi=0$ implies that 
$\Psi$ does not depend explicitly on $\bar{\theta}^1$ and $\bar{\theta}^2$, but implicitly through the combination
$y^\mu=x^\mu+i\theta_1\sigma^\mu\bar{\theta}^1+i\theta_2\sigma^\mu\bar{\theta}^2$ (cf. (\ref{chiral coordinate} and \cite{Lykken:1996xt})):
\begin{equation}
\Psi\equiv\Psi(y,\theta_1,\theta_2)\,.
\end{equation}
Expanding in powers of, say, $\theta_2$, one finds (e.g. \cite{Lykken:1996xt})
\begin{equation}
\Psi(y,\theta_1,\theta_2)=A(y,\theta_1)+\sqrt{2}\theta_2W(y,\theta_1)+(\theta_2)^2G(y,\theta_1)\,,
\end{equation}
where $A(y,\theta_1)$ and $G(y,\theta_1)$ are $\mathcal{N}=1$ scalar superfields and
$W^a(y,\theta_1)$ is an $\mathcal{N}=1$ spinor superfield.
However, for our purposes, we will consider the full expansion in $\theta_1$ and $\theta_2$:
\begin{eqnarray}
\Psi(X_{\alpha\beta},\theta_{1\alpha},\theta_{2\alpha},\varphi_{11},\varphi_{12},\varphi_{22})&=&
A+\sqrt{2}\theta_{1\alpha}\lambda^\alpha+\sqrt{2}\theta_{2\alpha}\chi^\alpha+\theta_{1\alpha}\theta_{2\beta}Z^{\alpha\beta}
+\theta_{1\alpha}\theta_{1\beta} V^{\alpha\beta}+\nonumber\\
&&\theta_{2\alpha}\theta_{2\beta} Y^{\alpha\beta}+\theta_{1\alpha}\theta_{1\beta}\theta_{2\gamma} E^{\alpha\beta\gamma}
+\theta_{2\alpha}\theta_{2\beta}\theta_{1\gamma} F^{\alpha\beta\gamma}+\nonumber\\
&&\theta_{1\alpha}\theta_{1\beta}\theta_{2\gamma}\theta_{2\delta} D^{\alpha\beta\gamma\delta}+\dots\,.
\end{eqnarray}
Here all the components are functions of $(X,\varphi)\equiv(X_{\alpha\beta},\varphi_{11},\varphi_{12},\varphi_{22})$ and the dots 
represent higher-order terms (cubic and quartic in $\theta_{1\alpha}$ and $\theta_{2\alpha}$ 
as well as products thereof) that will vanish on the light-cone.
The fields $V$, $Y$, $E$, $F$ and $D$ are anti-symmetric in $\alpha\beta$ (as well as in $\gamma\delta$ for $D$).
All the components transform as follows under a scaling transformation of the coordinates: 
$A(\lambda X,\lambda\varphi)=\lambda^{-\Delta}A(X,\varphi)$,
$\lambda^\alpha(\lambda X,\lambda\varphi)=\lambda^{-\Delta-1}\lambda^\alpha(X,\varphi)$ (same for $\chi^\alpha$),
$Z^{\alpha\beta}(\lambda X,\lambda\varphi)=\lambda^{-\Delta-2}Z^{\alpha\beta}(X,\varphi)$ 
(same for $V^{\alpha\beta}$ and $Y^{\alpha\beta}$), etc.
The four-dimensional $\mathcal{N}=2$ chiral superfield is defined on the light-cone as 
\begin{eqnarray}
\Psi(y,\theta_{1a},\theta_{2a})&\equiv&(X^+)^\Delta\Psi(X_{\alpha\beta},\theta_{1\alpha},\theta_{2\alpha},\varphi_{11},\varphi_{12},\varphi_{22})=\nonumber\\
&&A(y)+\sqrt{2}\theta_{1a}\lambda^a(y)+\sqrt{2}\theta_{2a}\chi^a(y)+\theta_{1a}\theta_{2b}Z^{ab}(y)+(\theta_1)^2 B(y)+\nonumber\\
&&(\theta_2)^2C(y)+(\theta_1)^2\theta_{2a}E^a(y)+(\theta_2)^2\theta_{1a}F^a(y)+(\theta_1)^2(\theta_2)^2D(y)\,.
\end{eqnarray}
In the following, we will be using such a notation for the components of this $\mathcal{N}=2$ off-shell multiplet.
There are a total of $9=3^2$ fields, of which:
\begin{itemize}
\item 1 spin-0 scalar field $A$
\item 2 spin-$\frac{1}{2}$ fields $\lambda$ and $\chi$
\item 1 spin-1 tensor field $Z$
\item 2 spin-1 fields $B$ and $C$
\item 2 spin-$\frac{3}{2}$ fields $E$ and $F$
\item 1 spin-2 field $D$.
\end{itemize}
This list agrees with the ones in e.g. \cite{Ketov:1988fy,Ketov:1988rw}.
These numbers make up the Pascal pyramid\footnote{In Appendix \ref{Appendix Pascal pyramid} we give all the necessary 
material about the Pascal pyramid. 
For the interested reader, more technical information can be found in \cite{combinatorics,Horn03}.} 
at layer 2:
\begin{equation}
\label{pascal pyramid for N=2 chiral superfield}
\begin{array}{ccccc}
1&&2&&1\\
&2&&2&\\
&&1&&
\end{array}
\qquad\Longleftrightarrow\qquad
\begin{array}{ccccc}
A&&\lambda,\chi&&Z\\
&B,C&&E,F&\\
&&D&&
\end{array}
\end{equation}
In Appendix \ref{Appendix Pascal pyramid} we consider the $\mathcal{N}=4$ case: there, the triangle 
analogue to (\ref{pascal pyramid for N=2 chiral superfield})
is larger, contains 15 entries and corresponds to the pyramid at layer 4.
More generically, the component fields of various kind in an $\mathcal{N}$-extended supermultiplets 
are equal in number to the elements of the Pascal pyramid at layer $\mathcal{N}$. 
The connection with the trinomial with power $\mathcal{N}$, which is at the origin of the pyramid, 
arises from the fact that in the product 
\begin{equation}
\theta_{1\alpha_1}^{x_1} \theta_{2\alpha_2}^{x_2} \dots \theta_{\mathcal{N}\alpha_{\mathcal{N}}}^{x_{\mathcal{N}}}
\end{equation}
the powers $x_i$ can only have three values, $x_i=0,1,2$, with $i=1,\dots,\mathcal{N}$.

Using standard techniques, we can now identify the component fields. We find:
\begin{subequations}
\label{N=2 components of chiral superfield}
\begin{eqnarray}
A(y)&=&
(X^+)^{\Delta}A(X_{\alpha\beta},0)\\
\lambda^a(y)&=&
(X^+)^{\Delta+1} \left[\lambda^a(X_{\alpha\beta},0)
-iy^\mu(\bar{\sigma}_\mu)^{\dot{a}a}\lambda_{\dot{a}}(X_{\alpha\beta},0)\right]\\
\chi^a(y)&=&
(X^+)^{\Delta+1} \left[\chi^a(X_{\alpha\beta},0)
-iy^\mu(\bar{\sigma}_\mu)^{\dot{a}a}\chi_{\dot{a}}(X_{\alpha\beta},0)\right]\\
\label{N=2 components of chiral superfield: Z-term}
Z^{ab}(y)&=&
i(X^+)^{\Delta+1}\epsilon^{ab} \left[-4\frac{\partial A}{\partial\varphi_{12}}(X_{\alpha\beta},0)
+X_{\alpha\beta}Z_{\rm aver}^{\alpha\beta}(X_{\alpha\beta},0)\right]\\
B(y)&=&
i(X^+)^{\Delta+1} \left[2\frac{\partial A}{\partial\varphi_{11}}(X_{\alpha\beta},0)
-X_{\alpha\beta}V^{\alpha\beta}(X_{\alpha\beta},0)\right]\\
C(y)&=&
i(X^+)^{\Delta+1} \left[2\frac{\partial A}{\partial\varphi_{22}}(X_{\alpha\beta},0)
-X_{\alpha\beta}Y^{\alpha\beta}(X_{\alpha\beta},0)\right]\\
E^a(y)&=&
i(X^+)^{\Delta+2} \left[2\sqrt{2}\frac{\partial\lambda^a}{\partial\varphi_{12}}(X_{\alpha\beta},0)
+2\sqrt{2}\frac{\partial\chi^a}{\partial\varphi_{11}}(X_{\alpha\beta},0)\right.\nonumber\\
&&\phantom{i(X^+)^{\Delta+2}}
-2\sqrt{2}iy^\mu(\bar{\sigma}_\mu)^{\dot{a}a}\frac{\partial\lambda_{\dot{a}}}{\partial\varphi_{12}}(X_{\alpha\beta},0)
-2\sqrt{2}iy^\mu(\bar{\sigma}_\mu)^{\dot{a}a}\frac{\partial\chi_{\dot{a}}}{\partial\varphi_{11}}(X_{\alpha\beta},0)\nonumber\\
&&\phantom{i(X^+)^{\Delta+2}}
\left.-X_{\alpha\beta}E^{\alpha\beta a}(X_{\alpha\beta},0)
-X_{\alpha\beta}iy^\mu(\bar{\sigma}_\mu)^{\dot{a}a}E^{\alpha\beta}_{\phantom{\alpha\beta}\dot{a}}(X_{\alpha\beta},0)\right]\\
F^a(y)&=&
i(X^+)^{\Delta+2} \left[2\sqrt{2}\frac{\partial\chi^a}{\partial\varphi_{12}}(X_{\alpha\beta},0)
+2\sqrt{2}\frac{\partial\lambda^a}{\partial\varphi_{22}}(X_{\alpha\beta},0)\right.\nonumber\\
&&\phantom{i(X^+)^{\Delta+2}}-2\sqrt{2}iy^\mu(\bar{\sigma}_\mu)^{\dot{a}a}\frac{\partial\chi_{\dot{a}}}{\partial\varphi_{12}}(X_{\alpha\beta},0)
-2\sqrt{2}iy^\mu(\bar{\sigma}_\mu)^{\dot{a}a}\frac{\partial\lambda_{\dot{a}}}{\partial\varphi_{11}}(X_{\alpha\beta},0)\nonumber\\
&&\phantom{i(X^+)^{\Delta+2}}\left.-X_{\alpha\beta}F^{\alpha\beta a}(X_{\alpha\beta},0)
-X_{\alpha\beta}iy^\mu(\bar{\sigma}_\mu)^{\dot{a}a}F^{\alpha\beta}_{\phantom{\alpha\beta}\dot{a}}(X_{\alpha\beta},0)\right]\\
\label{N=2 components of chiral superfield: D-term}
D(y)&=&
(X^+)^{\Delta+2} \left[2\frac{\partial^2 A}{\partial\varphi_{12}^2}(X_{\alpha\beta},0)
-2X_{\alpha\beta}\frac{\partial Z_{\rm aver}^{\alpha\beta}}{\partial\varphi_{12}}(X_{\alpha\beta},0)
-X_{\alpha\beta}X_{\gamma\delta}D^{\alpha\beta\gamma\delta}(X_{\alpha\beta},0)\right.\nonumber\\
&&\phantom{(X^+)^{\Delta+2}}\left.+2X_{\alpha\beta}\frac{\partial V^{\alpha\beta}}{\partial\varphi_{22}}(X_{\alpha\beta},0)
+2X_{\alpha\beta}\frac{\partial Y^{\alpha\beta}}{\partial\varphi_{11}}(X_{\alpha\beta},0)\right]\,.
\end{eqnarray}
\end{subequations}
In order to derive the components above, we have made use of the identity 
\begin{equation}
\label{light-cone identity}
\theta_{I\alpha}\theta_{I\beta} = -i X^+ (\theta_I)^2 X_{\alpha\beta}
\end{equation}
which is valid on the light-cone for each $I$. 
Note that a similar identity does not hold for $\theta_{I\alpha}\theta_{J\beta}$ if $I\neq J$.
For this reason, and in order to simplify the formulas above, we have introduced an auxiliary, or average, 
variable $Z^{\alpha\beta}_{\rm aver}$, defined as:
\begin{equation}
\label{Zaverage}
\theta_{1\alpha}\theta_{2\beta}Z^{\alpha\beta}(X,\varphi)
\equiv -i X^+(\theta_1\cdot\theta_2) X_{\alpha\beta}Z_{\rm aver}^{\alpha\beta}(X,\varphi)\,.
\end{equation}
The quantity $Z^{\alpha\beta}_{\rm aver}$ makes the scaling properties of the component explicit. 
It appears in (\ref{N=2 components of chiral superfield: Z-term}) and 
(\ref{N=2 components of chiral superfield: D-term}) and should be viewed as a shortcut of more lengthy expressions
corresponding to the expansion:
\begin{eqnarray}
\theta_{1\alpha}\theta_{2\beta}Z^{\alpha\beta}(X,\varphi)&=&
(X^+)^2\theta_{1a}\theta_{2b}
\left[Z^{ab}(X,\varphi)+iy^\mu(\bar{\sigma}_\mu)^{\dot{b}b}Z^{a}_{\phantom{a}\dot{b}}(X,\varphi)\right.\nonumber\\
&&
\left.+iy^\mu(\bar{\sigma}_\mu)^{\dot{a}a}Z_{\dot{a}}^{\phantom{\dot{a}}b}(X,\varphi)-
y^\mu y^\nu (\bar{\sigma}_\mu)^{\dot{a}a}(\bar{\sigma}_\nu)^{\dot{b}b}Z_{\dot{a}\dot{b}}(X,\varphi)\right]=\nonumber\\
&=&
(X^+)^2\theta_{1a}\theta_{2b}\left[Z^{ab}(X,\varphi)+iy^\mu(\bar{\sigma}_\mu)^{\dot{b}b}Z^{a}_{\phantom{a}\dot{b}}(X,\varphi)+
iy^\mu(\bar{\sigma}_\mu)^{\dot{a}a}Z_{\dot{a}}^{\phantom{\dot{a}}b}(X,\varphi)\right.\nonumber\\
&&
\left.+\frac{1}{2}\left(y^2\epsilon^{\dot{a}\dot{b}}\epsilon^{ab}-
4y_\mu y_\nu (\epsilon\sigma^{\lambda\mu})^{ab}(\bar{\sigma}^{\lambda\nu}\epsilon)^{\dot{a}\dot{b}}\right)
Z_{\dot{a}\dot{b}}(X,\varphi)\right]\,.
\end{eqnarray}
In going from the first equality to the second we have exploited the fact that the product $y^\mu y^\nu$ is 
symmetric in $\mu \nu$ and hence we can use the symmetric version of the product of two (barred) sigma matrices 
(see Appendix \ref{Appendix useful identities}).
The r.h.s. of (\ref{Zaverage}) is in components:
\begin{equation}
-iX^+\theta_{1}\cdot\theta_{2}
\big[
X_{ab}Z_{\rm aver}^{ab}(X,\varphi)
+X_a^{\phantom{a}\dot{b}}Z^{\phantom{{\rm aver}} a}_{{\rm aver}\phantom{a}\dot{b}}(X,\varphi)
+X^{\dot{a}}_{\phantom{\dot{a}}b}Z_{{\rm aver}\,\,\dot{a}}^{\phantom{{\rm aver}\,\,\dot{a}}b}(X,\varphi)
+X^{\dot{a}\dot{b}}Z_{{\rm aver}\,\,\dot{a}\dot{b}}(X,\varphi)
\big]
\end{equation}
and hence one can derive the components of $Z_{\rm aver}$, using (\ref{Coordinates of 4d superspace}) for the field $X_{\alpha\beta}$.
Expanding now both sides in $\varphi_{IJ}$, we are left with ordinary fields depending on $X_{\alpha\beta}$ only. We obtain:
\begin{eqnarray}
\theta_{1\alpha}\theta_{2\beta}Z^{\alpha\beta}(X,0)&=&
(X^+)^2\theta_{1a}\theta_{2b}\Big[Z^{ab}(X,0)\nonumber\\
&&
+iy^\mu(\bar{\sigma}_\mu)^{\dot{b}b}Z^{a}_{\phantom{a}\dot{b}}(X,0)
+iy^\mu(\bar{\sigma}_\mu)^{\dot{a}a}Z_{\dot{a}}^{\phantom{\dot{a}}b}(X,0)\nonumber\\
&&
+\frac{1}{2}\left(y^2\epsilon^{\dot{a}\dot{b}}\epsilon^{ab}-
4y_\mu y_\nu (\epsilon\sigma^{\lambda\mu})^{ab}(\bar{\sigma}^{\lambda\nu}\epsilon)^{\dot{a}\dot{b}}\right)
Z_{\dot{a}\dot{b}}(X,0)\Big]\\
\theta_{1\alpha}\theta_{2\beta}4iX^+\theta_1\cdot\theta_2\frac{\partial Z^{\alpha\beta}}{\partial\varphi_{12}}(X,0)&=&
(X^+)^2\theta_{1a}\theta_{2b}\Big[4iX^+\theta_1\cdot\theta_2\left(
\frac{\partial Z^{ab}}{\partial\varphi_{12}}(X,0)\right.\nonumber\\
&&
+iy^\mu(\bar{\sigma}_\mu)^{\dot{b}b}\frac{\partial Z^{a}_{\phantom{a}\dot{b}}}{\partial\varphi_{12}}(X,0)+
iy^\mu(\bar{\sigma}_\mu)^{\dot{a}a}\frac{Z_{\dot{a}}^{\phantom{\dot{a}}b}}{\partial\varphi_{12}}(X,0)\nonumber\\
&&
\left.+\frac{1}{2}\left(y^2\epsilon^{\dot{a}\dot{b}}\epsilon^{ab}-
4y_\mu y_\nu (\epsilon\sigma^{\lambda\mu})^{ab}(\bar{\sigma}^{\lambda\nu}\epsilon)^{\dot{a}\dot{b}}\right)
\frac{\partial Z_{\dot{a}\dot{b}}}{\partial\varphi_{12}}(X,0)
\right)
\Big]\,.
\end{eqnarray}
This modifies (\ref{N=2 components of chiral superfield: Z-term}) and (\ref{N=2 components of chiral superfield: D-term}),
which explicitly read:
\begin{eqnarray}
Z^{ab}(y)&=&
(X^+)^{\Delta+1}
\left[-4i\epsilon^{ab}\frac{\partial A}{\partial\varphi_{12}}(X,0)\right.\nonumber\\
&&
\phantom{(X^+)^{\Delta+1}}
+X^+\Big[Z^{ab}(X,0)
+iy^\mu(\bar{\sigma}_\mu)^{\dot{b}b}Z^{a}_{\phantom{a}\dot{b}}(X,0)
+iy^\mu(\bar{\sigma}_\mu)^{\dot{a}a}Z_{\dot{a}}^{\phantom{\dot{a}}b}(X,0)\nonumber\\
&&
\phantom{(X^+)^{\Delta+1}}
\phantom{(X^+)}
\left.+\frac{1}{2}\left(y^2\epsilon^{\dot{a}\dot{b}}\epsilon^{ab}-
4y_\mu y_\nu (\epsilon\sigma^{\lambda\mu})^{ab}(\bar{\sigma}^{\lambda\nu}\epsilon)^{\dot{a}\dot{b}}\right)
Z_{\dot{a}\dot{b}}(X,0)\Big]\right]\\
D(y)&=&
(X^+)^{\Delta+2} \left[2\frac{\partial^2 A}{\partial\varphi_{12}^2}(X,0)
-X_{\alpha\beta}X_{\gamma\delta}D^{\alpha\beta\gamma\delta}(X,0)
+2X_{\alpha\beta}\frac{\partial V^{\alpha\beta}}{\partial\varphi_{22}}(X,0)
+2X_{\alpha\beta}\frac{\partial Y^{\alpha\beta}}{\partial\varphi_{11}}(X,0)
\right.\nonumber\\
&&
\phantom{(X^+)^{\Delta+2}}
\left.
-i\epsilon_{ab}X^+\Big[
\frac{\partial Z^{ab}}{\partial\varphi_{12}}(X,0)
+iy^\mu(\bar{\sigma}_\mu)^{\dot{b}b}\frac{\partial Z^{a}_{\phantom{a}\dot{b}}}{\partial\varphi_{12}}(X,0)+
iy^\mu(\bar{\sigma}_\mu)^{\dot{a}a}\frac{Z_{\dot{a}}^{\phantom{\dot{a}}b}}{\partial\varphi_{12}}(X,0)
\right.\nonumber\\
&&
\phantom{(X^+)^{\Delta+2}}
\phantom{-i\epsilon_{ab}(X^+)}
\left.+\frac{1}{2}\left(y^2\epsilon^{\dot{a}\dot{b}}\epsilon^{ab}-
4y_\mu y_\nu (\epsilon\sigma^{\lambda\mu})^{ab}(\bar{\sigma}^{\lambda\nu}\epsilon)^{\dot{a}\dot{b}}\right)
\frac{\partial Z_{\dot{a}\dot{b}}}{\partial\varphi_{12}}(X,0)
\Big]
\right]\,.
\end{eqnarray}

Using (\ref{component transformations from superfield}) we can determine the superconformal transformations for the 
components of the $\mathcal{N}=2$ superfield. We obtain:
\begin{subequations}
\begin{eqnarray}
\delta A &=&
\sqrt{2}\Big(\epsilon_1-iy_\mu\bar{\eta}^1\bar{\sigma}^\mu\Big)\cdot\lambda + 
\sqrt{2}\Big(\epsilon_2-iy_\mu\bar{\eta}^2\bar{\sigma}^\mu\Big)\cdot\chi\\
\delta \lambda^a &=&
2\sqrt{2}\Delta\eta^{1a}A + \sqrt{2}\epsilon^{ab}
\Big(i(\sigma^\mu)_{b\dot{a}}\bar{\epsilon}^{1\dot{a}}-y_\nu(\sigma^\mu)_{b\dot{a}}(\bar{\sigma}^\nu)^{\dot{a}c}\eta^1_c\Big)
\partial_\mu A\nonumber\\
&&+\sqrt{2}\Big(\epsilon_1^a-iy_\mu(\bar{\sigma}^\mu)^{\dot{a}a}\bar{\eta}^1_{\dot{a}}\Big)B +
\frac{\sqrt{2}}{2}\Big(\epsilon_2^c-iy_\mu(\bar{\sigma}^\mu)^{\dot{a}c}\bar{\eta}^2_{\dot{a}}\Big)\epsilon_{cb}Z^{ab}\\
\delta \chi^a &=&
2\sqrt{2}\Delta\eta^{2a}A + \sqrt{2}\epsilon^{ab}
\Big(i(\sigma^\mu)_{b\dot{a}}\bar{\epsilon}^{2\dot{a}}-y_\nu(\sigma^\mu)_{b\dot{a}}(\bar{\sigma}^\nu)^{\dot{a}c}\eta^2_c\Big)
\partial_\mu A\nonumber\\
&&+\sqrt{2}\Big(\epsilon_2^a-iy_\mu(\bar{\sigma}^\mu)^{\dot{a}a}\bar{\eta}^2_{\dot{a}}\Big)C -
\frac{\sqrt{2}}{2}\Big(\epsilon_1^c-iy_\mu(\bar{\sigma}^\mu)^{\dot{a}c}\bar{\eta}^1_{\dot{a}}\Big)\epsilon_{cb}Z^{ba}\\
\delta Z^{ab} &=&
-4\sqrt{2}\Delta\eta^{1a}\chi^b-4\sqrt{2}\Delta\lambda^a\eta^{2b} + \sqrt{2}\epsilon^{ac}
\Big(2i(\sigma^\mu)_{c\dot{a}}\bar{\epsilon}^{1\dot{a}}-2y_\nu(\sigma^\mu)_{c\dot{a}}(\bar{\sigma}^\nu)^{\dot{a}d}\eta^1_d\Big)\partial_\mu\chi^b
\nonumber\\
&&-\sqrt{2}\epsilon^{bc}
\Big(2i(\sigma^\mu)_{c\dot{a}}\bar{\epsilon}^{2\dot{a}}-2y_\nu(\sigma^\mu)_{c\dot{a}}(\bar{\sigma}^\nu)^{\dot{a}d}\eta^2_d\Big)\partial_\mu\lambda^a
-2\Big(\epsilon_1^a-iy_\mu(\bar{\sigma}^\mu)^{\dot{a}a}\bar{\eta}^1_{\dot{a}}\Big)E^b\nonumber\\
&&+2\Big(\epsilon_2^a-iy_\mu(\bar{\sigma}^\mu)^{\dot{a}a}\bar{\eta}^2_{\dot{a}}\Big)F^b
-4\sqrt{2}\epsilon^{ab}\eta^2\cdot\lambda - 4\sqrt{2}\epsilon^{ab}\eta^1\cdot\chi\\
\delta B &=&
2\sqrt{2}\Delta\eta^1\cdot\lambda
+\sqrt{2}\Big(i(\sigma^\mu)_{b\dot{a}}\bar{\epsilon}^{1\dot{a}}-y_\nu(\sigma^\mu)_{b\dot{a}}(\bar{\sigma}^\nu)^{\dot{a}c}\eta^1_c\Big)
\partial_\mu \lambda^b + \Big(\epsilon_2^a-iy_\mu(\bar{\sigma}^\mu)^{\dot{a}a}\bar{\eta}^2_{\dot{a}}\Big)E_a\\
\delta C &=&
2\sqrt{2}\Delta\eta^2\cdot\chi
+\sqrt{2}\Big(i(\sigma^\mu)_{b\dot{a}}\bar{\epsilon}^{2\dot{a}}-y_\nu(\sigma^\mu)_{b\dot{a}}(\bar{\sigma}^\nu)^{\dot{a}c}\eta^2_c\Big)
\partial_\mu \chi^b + \Big(\epsilon_1^a-iy_\mu(\bar{\sigma}^\mu)^{\dot{a}a}\bar{\eta}^1_{\dot{a}}\Big)F_a\\
\delta E^a &=&
(4-2\Delta)\eta^1_b Z^{ba} - 2 Z^{ab} \eta^1_b + (4\Delta+4)\eta^{2a}B
-\Big(i(\sigma^\mu)_{b\dot{a}}\bar{\epsilon}^{1\dot{a}}-y_\nu(\sigma^\mu)_{b\dot{a}}(\bar{\sigma}^\nu)^{\dot{a}c}\eta^1_c\Big)
\partial_\mu Z^{ba} \nonumber\\
&&+\epsilon^{ab}
\Big(2i(\sigma^\mu)_{b\dot{a}}\bar{\epsilon}^{2\dot{a}}-2y_\nu(\sigma^\mu)_{b\dot{a}}(\bar{\sigma}^\nu)^{\dot{a}c}\eta^2_c\Big)
\partial_\mu B + 2 \Big(\epsilon_2^a-iy_\mu(\bar{\sigma}^\mu)^{\dot{a}a}\bar{\eta}^2_{\dot{a}}\Big) D\\
\delta F^a &=&
2\eta^2_b Z^{ba} - (4-2\Delta) Z^{ab} \eta^2_b + (4\Delta+4) \eta^{1a}C
+\Big(i(\sigma^\mu)_{b\dot{a}}\bar{\epsilon}^{2\dot{a}}-y_\nu(\sigma^\mu)_{b\dot{a}}(\bar{\sigma}^\nu)^{\dot{a}c}\eta^2_c\Big)
\partial_\mu Z^{ba} \nonumber\\
&&+\epsilon^{ab}
\Big(2i(\sigma^\mu)_{b\dot{a}}\bar{\epsilon}^{1\dot{a}}-2y_\nu(\sigma^\mu)_{b\dot{a}}(\bar{\sigma}^\nu)^{\dot{a}c}\eta^1_c\Big)
\partial_\mu C + 2 \Big(\epsilon_1^a-iy_\mu(\bar{\sigma}^\mu)^{\dot{a}a}\bar{\eta}^1_{\dot{a}}\Big) D \\
\delta D &=&
(2-2\Delta)\eta^1\cdot F + (2-2\Delta)\eta^2\cdot E
-\Big(i(\sigma^\mu)_{b\dot{a}}\bar{\epsilon}^{1\dot{a}}-y_\nu(\sigma^\mu)_{b\dot{a}}(\bar{\sigma}^\nu)^{\dot{a}c}\eta^1_c\Big)
\partial_\mu F^b\nonumber\\
&&-\Big(i(\sigma^\mu)_{b\dot{a}}\bar{\epsilon}^{2\dot{a}}-y_\nu(\sigma^\mu)_{b\dot{a}}(\bar{\sigma}^\nu)^{\dot{a}c}\eta^2_c\Big)
\partial_\mu E^b\,.
\end{eqnarray}
\end{subequations}
Special conformal transformations correspond to $\epsilon_I^a=0$, as it was written in \cite{Goldberger:2011yp}.
By setting $\epsilon_I^a=0$, we see that at the origin the $\theta_I=\bar{\theta}_I=0$ component is invariant,
namely $\delta A(0)=0$, and hence $A(0)$ is annihilated by the special superconformal generators $S^I_a,\bar{S}_I^{\dot{a}}$,
defined by 
\begin{equation}
Q^I_\alpha=\frac{i}{2}\left(
\begin{array}{c}
-Q^I_a \\ \bar{S}_I^{\dot{a}}
\end{array}
\right)
\qquad{\rm and}\qquad
\bar{Q}_I^\alpha=\frac{i}{2}\left(
\begin{array}{c}
-S_I^a \\ \bar{Q}^I_{\dot{a}}
\end{array}
\right)\,.
\end{equation}
This can be checked by using the appropriate generalizations of those operators from the $\mathcal{N}=1$ case of \cite{Goldberger:2011yp},
which tells us that $\Phi(y,\theta_I)$ is a chiral field generated by a chiral primary operator $A$.
This is a generic property of the chiral sector. In fact, using the transformation rules (\ref{4d space-time superconformal transformations}), 
for special conformal transformations, $\delta A\equiv\delta\Phi|=-\delta\theta_I^a\frac{\partial\Phi}{\partial\theta_I^a}|$ 
(the vertical bar denoting the $\theta_I=0$ component of the r.h.s., which corresponds to taking the spinorial $\theta$-component of $\Phi$) 
will always be proportional to $y^\mu$ and hence will vanish at the origin.

\section{Conclusion}
In this paper we generalized the construction of superembedding methods for 4d $\mathcal{N}=1$ in \cite{Goldberger:2011yp}
to embeddings of $\mathcal{N}$-extended superconformal field theories.
In this way, the superconformal group acts linearly on the coordinates of the ambient space,
the conformal symmetry is manifest and moreover the method is valid for any conformally flat space, not just Minkowski.
We have considered explicitly the case of $\mathcal{N}=2$ chiral superfields in four dimensions 
and concluded that any conformal chiral superfield of the $\mathcal{N}$-extended supersymmetry is generated
by a chiral primary operator sitting in its highest component.
We have also noted a correspondence 
between the number of component fields of a certain type in any chiral multiplet of the $\mathcal{N}$-extended supersymmetry 
and the entries of the Pascal pyramid at layer $\mathcal{N}$.  
This correspondence has been explicitly showed in the cases of $\mathcal{N}=2$ and $\mathcal{N}=4$ supersymmetry.

We have not considered correlation functions, that for $\mathcal{N}=1$ were presented in \cite{Goldberger:2011yp}, 
neither non-holomorphic operators and higher-rank tensors, which among other things are relevant in AdS/CFT and its applications. 
These points are left as open questions.

\section*{Acknowledgements}
We would like to acknowledge Beatriz Gato Rivera for carefully reading the manuscript,
the Instituto de Fisica Teorica in Madrid, IFT-CSIC/UAM, for hospitality during the completion of this work, 
Warren Siegel, Andreas Stergiou and Dimitri Sorokin for correspondence, 
and the NPB referee for useful comments. 
This research has been supported by funding of the Project CONSOLIDER-INGENIO 2010, Program CPAN (CSD2007-00042).

\section*{Appendix}

\appendix

\section{Explicit SU(2,2) transformations of \texorpdfstring{$X_{\alpha\beta}$}{} }
\label{explicit SU(2,2) transformations}
For the six-dimensional vector $X^m=(X^+,X^\mu,X^-)$ the transformations under the conformal group $SO(4,2)$ are $\delta X^m=\omega^{mn}X_n$,
with $\omega^{mn}$ anti-symmetric. The four-dimensional coordinates $x^\mu$ are related to the six-dimensional ones by solving 
the light-cone constraint:
\begin{equation}
X^+= {\rm constant} \,,\quad x^\mu=\frac{X^\mu}{X^+} \,,\quad X^-=- X^+ x^2\,.
\end{equation}
Using the conventions of \cite{Goldberger:2011yp}, since\footnote{The index structure of the tensors 
$X_{\alpha\beta}$ and $\bar{X}^{\alpha\beta}$ is
\begin{equation}
X_{\alpha\beta}=\left(
\begin{array}{cc}
X_{ab} & X_a^{\phantom{a}\dot{b}}\\
X^{\dot{a}}_{\phantom{\dot{a}}b} & X^{\dot{a}\dot{b}}
\end{array}
\right)
\qquad {\rm and} \qquad 
\bar{X}^{\alpha\beta}=\left(
\begin{array}{cc}
\bar{X}^{ab} & \bar{X}^a_{\phantom{a}\dot{b}}\\
\bar{X}_{\dot{a}}^{\phantom{\dot{a}}b} & \bar{X}_{\dot{a}\dot{b}}
\end{array}
\right)\,.
\end{equation}
}
$X_{\alpha\beta}=\frac{1}{2}X_m\tilde{\Gamma}^m_{\alpha\beta}$ in $SU(2,2)$ notation, these transformations are:
\begin{eqnarray}
\label{Generic SU(2,2) transformation}
\delta X_{\alpha\beta}&=&\frac{1}{2}\delta X_m\tilde{\Gamma}^m_{\alpha\beta}=
\frac{1}{2}\delta X_+\tilde{\Gamma}^+_{\alpha\beta}+
\frac{1}{2}\delta X_\mu\tilde{\Gamma}^\mu_{\alpha\beta}+
\frac{1}{2}\delta X_-\tilde{\Gamma}^-_{\alpha\beta}=\nonumber\\
&=&
\frac{1}{4}\delta X^-\tilde{\Gamma}^+_{\alpha\beta}+
\frac{1}{2}\delta X^\nu\eta_{\mu\nu}\tilde{\Gamma}^\mu_{\alpha\beta}+
\frac{1}{4}\delta X^+\tilde{\Gamma}^-_{\alpha\beta}=\nonumber\\
&=&\left(
\begin{array}{cc}
i\frac{1}{2}\delta X^+\epsilon_{ab} & \frac{1}{2}\eta_{\mu\nu}\delta X^\nu (\sigma^\mu)_{a\dot{d}}\epsilon^{\dot{d}\dot{b}}\\
-\frac{1}{2}\eta_{\mu\nu}\delta X^\nu (\bar{\sigma}^\mu)^{\dot{a}d}\epsilon_{db} & i\frac{1}{2}\delta X^-\epsilon^{\dot{a}\dot{b}}
\end{array}
\right)\,.
\end{eqnarray}
In the next subsections, we consider the single conformal transformations explicitly.
We will also apply these transformations to the origin (\ref{origin in spinor notation}) and will find
the summarizing table \ref{transformations of the origin}, that is given here for convenience.
\begin{table}[ht]
\begin{center}
\caption{Summary of $SU(2,2)$ transformations of the origin}
\begin{tabular}{|c|c|}
\hline
&\\
Translations & 
$\delta \hat{X}_{\alpha\beta}=\frac{1}{4}\eta_{\mu\nu}\omega^{\nu-}X^+\cdot
\left(
\begin{array}{cc}
0 & (\sigma^\mu)_{a\dot{d}}\epsilon^{\dot{d}\dot{b}}\\
-(\bar{\sigma}^\mu)^{\dot{a}d}\epsilon_{db} & 0
\end{array}
\right)$\\
\hline
&\\
Lorentz & 
$\delta \hat{X}_{\alpha\beta}=0$\\
\hline
&\\
Special Conformal & 
$\delta \hat{X}_{\alpha\beta}=0$\\
\hline
&\\
Dilatations & 
$\delta \hat{X}_{\alpha\beta}=\frac{1}{4}\omega^{+-}\cdot
\left(
\begin{array}{cc}
i\epsilon_{ab} X^+ & 0\\
0 & 0
\end{array}
\right)$\\
\hline
\end{tabular}
\label{transformations of the origin}
\end{center}
\end{table}

\subsection{Translations}
The only non-zero parameter is $\omega^{\mu -}$. Hence
\begin{subequations}
\begin{eqnarray}
\delta X^+ &=& \omega^{+ n}X_n=0\\
\delta X^\mu &=& \omega^{\mu n}X_n=\omega^{\mu -}X_-=\frac{1}{2}\omega^{\mu-}X^+\\
\delta X^- &=& \omega^{- n}X_n=-\omega^{\mu-}X_\mu=-\eta_{\mu\nu}\omega^{\mu-}X^\nu\,.
\end{eqnarray}
\end{subequations}
The second line gives the four-dimensional translation
\begin{equation}
\delta x^\mu=\frac{1}{2}\omega^{\mu-}
\end{equation}
and consistently the third line gives $\delta x^2=\eta_{\mu\nu}\omega^{\mu-}x^\nu$.
Using (\ref{Generic SU(2,2) transformation}) one has:
\begin{equation}
\delta X_{\alpha\beta}=\frac{1}{2}\eta_{\mu\nu}\omega^{\nu-}\cdot
\left(
\begin{array}{cc}
0 & \frac{1}{2} X^+(\sigma^\mu)_{a\dot{d}}\epsilon^{\dot{d}\dot{b}}\\
-\frac{1}{2} X^+(\bar{\sigma}^\mu)^{\dot{a}d}\epsilon_{db} & -iX^\mu
\end{array}
\right)\,.
\end{equation}
When evaluated at the origin, this becomes
\begin{equation}
\delta \hat{X}_{\alpha\beta}=\frac{1}{4}\eta_{\mu\nu}\omega^{\nu-}X^+\cdot
\left(
\begin{array}{cc}
0 & (\sigma^\mu)_{a\dot{d}}\epsilon^{\dot{d}\dot{b}}\\
-(\bar{\sigma}^\mu)^{\dot{a}d}\epsilon_{db} & 0
\end{array}
\right)\,.
\end{equation}

\subsection{Lorentz transformations}
The only non-zero parameter is $\omega^{\mu\nu}$. This implies:
\begin{subequations}
\begin{eqnarray}
\delta X^+ &=& \omega^{+ n}X_n=0\\
\delta X^\mu &=& \omega^{\mu\nu}X_\nu=\omega^{\mu\nu}\eta_{\nu\rho}X^{\rho}\\
\delta X^- &=& \omega^{- n}X_n=0\,.
\end{eqnarray}
\end{subequations}
The second line implies the four-dimensional Lorentz transformation
\begin{equation}
\delta x^\mu=\omega^{\mu\nu}x_\nu
\end{equation}
and consistently the third line gives $\delta x^2=0$. In $SU(2,2)$ notation, one has
\begin{equation}
\delta X_{\alpha\beta}=\frac{1}{2}\eta_{\mu\nu}\omega^{\nu\rho}X_\rho\cdot
\left(
\begin{array}{cc}
0 & (\sigma^\mu)_{a\dot{d}}\epsilon^{\dot{d}\dot{b}}\\
-(\bar{\sigma}^\mu)^{\dot{a}d}\epsilon_{db} & 0
\end{array}
\right)\,,
\end{equation}
which vanishes at the origin,
\begin{equation}
\delta \hat{X}_{\alpha\beta}=0\,.
\end{equation}

\subsection{Special conformal transformations}
Here $\omega^{\mu+}$ is non-zero and
\begin{subequations}
\begin{eqnarray}
\delta X^+ &=& \omega^{+ n}X_n=\omega^{+\mu}X_\mu=-\omega^{\mu+}X_\mu\\
\delta X^\mu &=& \omega^{\mu n}X_n=\omega^{\mu+}X_+=\frac{1}{2}\omega^{\mu+}X^-\\
\delta X^- &=& \omega^{- n}X_n=0\,.
\end{eqnarray}
\end{subequations}
The first two lines combined give the four-dimensional special conformal transformation
\begin{equation}
\delta x^\mu=x^2 b^\mu-2x^\mu(b\cdot x)\,,
\end{equation}
with $b^\mu=-\frac{1}{2}\omega^{\mu+}$.
In $SU(2,2)$ notation, one has
\begin{equation}
\delta X_{\alpha\beta}=\frac{1}{2}\eta_{\mu\nu}\omega^{\nu+}\cdot
\left(
\begin{array}{cc}
-iX_\mu & \frac{1}{2}X^-(\sigma^\mu)_{a\dot{d}}\epsilon^{\dot{d}\dot{b}}\\
-\frac{1}{2}X^-(\bar{\sigma}^\mu)^{\dot{a}d}\epsilon_{db} & 0
\end{array}
\right)\,,
\end{equation}
which vanishes at the origin,
\begin{equation}
\delta \hat{X}_{\alpha\beta}=0\,.
\end{equation}

\subsection{Dilatations}
The non-vanishing parameter is $\omega^{+-}$ and
\begin{subequations}
\begin{eqnarray}
\delta X^+ &=& \omega^{+ n}X_n=\omega^{+-}X_-=\frac{1}{2}\omega^{+-}X^+\\
\delta X^\mu &=& \omega^{\mu n}X_n=0\\
\delta X^- &=& \omega^{- n}X_n=\omega^{-+}X_+=-\frac{1}{2}\omega^{+-}X^-\,.
\end{eqnarray}
\end{subequations}
The first two lines combined give the four-dimensional scale transformation
\begin{equation}
\delta x^\mu=-\frac{1}{2}\omega^{+-}x^\mu\,,
\end{equation}
while the third line gives consistently $\delta x^2=-\omega^{+-}x^2$.
In spinor notation,
\begin{equation}
\delta X_{\alpha\beta}=\frac{1}{4}\omega^{+-}\cdot
\left(
\begin{array}{cc}
iX^+\epsilon_{ab} & 0\\
0 & iX^-\epsilon^{\dot{a}\dot{b}}
\end{array}
\right)\,.
\end{equation}
At the origin this becomes
\begin{equation}
\delta \hat{X}_{\alpha\beta}=\frac{1}{4}\omega^{+-}\cdot
\left(
\begin{array}{cc}
iX^+\epsilon_{ab} & 0\\
0 & 0
\end{array}
\right)\,.
\end{equation}

\section{Notation and identities}
\label{Appendix useful identities}
In this appendix we write down some spinor identities that are used in the calculations of the main text. 
Our conventions for the spinors are the same as in \cite{Goldberger:2011yp}, in particular:
\begin{equation}
V_\alpha=\left(
\begin{array}{c}
\psi_a\\
\bar{\chi}^{\dot{a}}
\end{array}
\right)
\qquad \Longleftrightarrow \qquad
\bar{V}^\alpha=\bar{V}_{\dot{\alpha}}A^{\dot{\alpha}\alpha}=(\chi^a,\bar{\psi}_{\dot{a}})\,,
\end{equation}
where
\begin{equation}
A^{\dot{\alpha}\beta}=\left(
\begin{array}{cc}
0&\delta^{\dot{a}}_{\phantom{a}\dot{b}}\\
\delta_a^{\phantom{a}b}&0
\end{array}
\right)\,.
\end{equation}
The gamma matrices are given by:
\begin{equation}
\Gamma^{+\alpha\beta}=\left(
\begin{array}{cc}
2i\epsilon^{ab}&0\\
0&0
\end{array}
\right)
\,,\qquad
\Gamma^{-\alpha\beta}=\left(
\begin{array}{cc}
0&0\\
0&2i\epsilon_{\dot{a}\dot{b}}
\end{array}
\right)
\,,\qquad
\Gamma^{\mu\alpha\beta}=\left(
\begin{array}{cc}
0&\bar{\sigma}^{\mu\dot{d}a}\epsilon_{\dot{d}\dot{b}}\\
-\sigma^\mu_{\phantom{\mu}d\dot{a}}\epsilon^{db}&0
\end{array}
\right)
\end{equation}
and
\begin{equation}
\tilde{\Gamma}^+_{\alpha\beta}=\left(
\begin{array}{cc}
0&0\\
0&2i\epsilon^{\dot{a}\dot{b}}
\end{array}
\right)
\,,\qquad
\tilde{\Gamma}^-_{\alpha\beta}=\left(
\begin{array}{cc}
2i\epsilon_{ab}&0\\
0&0
\end{array}
\right)
\,,\qquad
\tilde{\Gamma}^\mu_{\alpha\beta}=\left(
\begin{array}{cc}
0&\sigma^\mu_{\phantom{\mu}a\dot{d}}\epsilon^{\dot{d}\dot{b}}\\
-\bar{\sigma}^{\mu\dot{a}d}\epsilon_{db}&0
\end{array}
\right)
\end{equation}
The $SO(4,2)$ coordinate $X^m$ and the anti-symmetric $SU(2,2)$ tensor $X_{\alpha\beta}$ are related by
\begin{equation}
X^m=\frac{1}{2}X_{\alpha\beta}\Gamma^{m\alpha\beta}=\frac{1}{2}X^{\alpha\beta}\tilde{\Gamma}^m_{\alpha\beta}
\end{equation}
and
\begin{equation}
X_{\alpha\beta}=\frac{1}{2}X_m\tilde{\Gamma}^m_{\alpha\beta}\,,
\qquad
X^{\alpha\beta}=\frac{1}{2}X_m\Gamma^{m\alpha\beta}\,.
\end{equation}

As it is standard in supersymmetry, for each $I$, the product $\bar{\theta}^{I\dot{a}}\theta_I^{b}$ 
must be proportional to $(\bar{\sigma^\mu})^{\dot{a}b}$, i.e.
$2\bar{\theta}^{I\dot{a}}\theta_I^{b}=A^\mu(\bar{\sigma}_\mu)^{\dot{a}b}$, 
where the proportionality constant $A^\mu$ is fixed by multiplying both sides by $(\sigma^\nu)_{b\dot{a}}$ and summing over $b$ and $\dot{a}$ 
(which gives ${\rm Tr}(\bar{\sigma}^\mu\sigma^\nu)=-2\eta^{\mu\nu}$).
After anti-commuting the theta's, one finds $A^\mu=\theta_I\sigma^\mu\bar{\theta}^I$ and hence, for each $I$:
\begin{equation}
\bar{\theta}^{I\dot{a}}\theta_I^{b}=
\frac{1}{2}\theta_I\sigma^\mu\bar{\theta}^I(\bar{\sigma}_\mu)^{\dot{a}b}\,.
\end{equation} 
This expression was used in (\ref{theta factor in the product}).
One can make similar manipulations in extended supersymmetry and write (repeated indices are not summed):
\begin{eqnarray}
\theta_{Ia}\theta_{Ib}&=&\frac{1}{2}(\theta_I)^2\epsilon_{ab}\\
\theta_I\cdot\theta_J&=&-\epsilon^{ab}\theta_{Ia}\theta_{Jb}\\
\theta_{Ia} (\theta_I\cdot\theta_J) &=& \frac{1}{2} (\theta_I)^2 \theta_{Ja}\\
(\theta_I\cdot\theta_J)^2 &=& -\frac{1}{2} (\theta_I)^2 (\theta_J)^2\\
\theta_{Ia}\theta_{Jb} (\theta_I\cdot\theta_J) &=& -\frac{1}{4} (\theta_I)^2 (\theta_J)^2\epsilon_{ab}\,,
\end{eqnarray}
which were used, for example, in (\ref{N=2 components of chiral superfield}) with $I,J=1,2$.

Our conventions for the sigma matrices are as in \cite{Goldberger:2011yp} and \cite{WB}.
In particular, $\sigma^\mu=\{-1,\sigma^i\}$ and $\bar{\sigma}^\mu=\{-1,-\sigma^i\}$.
These are related by
\begin{equation}
(\bar{\sigma}^\mu)^{\dot{a}a}=\epsilon^{\dot{a}\dot{b}}\epsilon^{ab}(\sigma^\mu)_{b\dot{b}}\,.
\end{equation}
This implies that 
\begin{equation}
(\sigma^\mu)_{a\dot{d}}\epsilon^{\dot{d}\dot{b}}=(\bar{\sigma}^\mu)^{\dot{b}d}\epsilon_{da}
\qquad
{\rm and}
\qquad
(\sigma^\mu)_{a\dot{a}}=\epsilon_{ad}\epsilon_{\dot{a}\dot{d}}(\bar{\sigma}^\mu)^{\dot{d}d}\,.
\end{equation}
The former equation was used in (\ref{Coordinates of 4d superspace}). There, we also used the following relation:
\begin{equation}
y^\mu y^\nu(\bar{\sigma}_\mu)^{\dot{a}c}(\bar{\sigma}_\nu)^{\dot{b}d}\epsilon_{cd}=y^\mu y^\nu\eta_{\mu\nu}\epsilon^{\dot{a}\dot{b}}\,.
\end{equation}
This comes from the more generic relation regarding the product of two sigma matrices, which can be expressed
in terms of the generators $(\sigma^{mn})_a^{\phantom{a}b}$ and $(\bar{\sigma}^{mn})^{\dot{a}}_{\phantom{\dot{a}}\dot{b}}$ 
of the Lorentz group in the spinor representation, which are anti-symmetric in the indices $m$ and $n$ (see below).
However, the symmetric part in the space-time indices and the anti-symmetric part in the spinorial indices
is fully specified by the metric tensor and the epsilon tensor as:
\begin{equation}
\bar{\sigma}_{(\mu}^{\phantom{\mu}\dot{a}[c}\bar{\sigma}_{\nu)}^{\phantom{\nu}|\dot{b}|d]}=
-\frac{1}{2}\eta_{\mu\nu}\epsilon^{\dot{a}\dot{b}}\epsilon^{cd}\,.
\end{equation}
A similar formula holds for the anti-symmetric combination of the dotted indices as well as 
for the product of two (unbarred) sigma matrices.

Using the relations above, it is straightforward to check that:
\begin{equation}
\bar{\epsilon}\bar{\sigma}^\mu\theta=-\theta\sigma^\mu\bar{\epsilon}
\qquad {\rm and} \qquad
\eta\sigma^\nu\bar{\sigma}^\mu\theta=\theta\sigma^\mu\bar{\sigma}^\nu\eta\,.
\end{equation}
These identities were used in (\ref{4d space-time superconformal transformations}).
If we only look at the symmetric part in the space-time indices, instead, we have to consider the more general expression
\begin{equation}
\bar{\sigma}_{(\mu}^{\phantom{\mu}\dot{a}a}\bar{\sigma}_{\nu)}^{\phantom{\nu}\dot{b}b}=
\frac{1}{2}[-\eta_{\mu\nu}\epsilon^{\dot{a}\dot{b}}\epsilon^{ab}+
4(\epsilon\sigma^{\lambda\mu})^{ab}(\bar{\sigma}^{\lambda\nu}\epsilon)^{\dot{a}\dot{b}}]\,,
\end{equation}
as it appears in \cite{WB}, which was used in (\ref{N=2 components of chiral superfield}).

\section{Barred 4D superspace}
\label{4d superspace barred}
In the translations given in (\ref{superconformal transformations}) 
we need to know the inverse of (\ref{explicit full transformation in 4d superspace}).
Because of (\ref{full transformation in 4d superspace}), the inverse is a product of the two commuting transformations
\begin{equation}
\label{full transformation in 4d superspace, inverse}
\mathcal{U}^{-1}(x^\mu,\{\theta_{Ia}\},\{\bar{\theta}^{J\dot{a}}\})_A^{\phantom{A}B}=
\mathcal{U}^{-1}(x^\mu,0,0)_A^{\phantom{A}C} 
\mathcal{U}^{-1}(0,\{\theta_{Ia}\},\{\bar{\theta}^{J\dot{a}}\})_C^{\phantom{C}B}\,.
\end{equation}
The space-time part is simply obtained by replacing $x^\mu\rightarrow-x^\mu$ in (\ref{space-time factor in the product}):
\begin{equation}
\label{space-time factor in the product, inverse}
\mathcal{U}^{-1}(x^\mu,0,0)_A^{\phantom{A}B}=
\left(
\begin{array}{cc}
\left(
e^{+i x_\mu \Sigma^{\mu +}}
\right)_\alpha^{\phantom{\alpha}\beta}
& 0\\
0 & \delta_I^{\phantom{I}J}
\end{array}
\right)=
\left(
\begin{array}{ccc}
\delta_a^{\phantom{a}b}&0&0\\
-ix^\mu(\bar{\sigma}_\mu)^{\dot{a}b}&\delta^{\dot{a}}_{\phantom{\dot{a}}\dot{b}}&0\\
0&0&\delta_I^{\phantom{I}J}
\end{array}
\right)\,.
\end{equation}
To compute the theta part, it is convenient to use the following expression for the inversion formula of block matrices:
\begin{equation}
\left(
\begin{array}{cc}
A&B\\
C&D
\end{array}
\right)^{-1}=
\left(
\begin{array}{cc}
(A-BD^{-1}C)^{-1}&-(A-BD^{-1}C)^{-1}BD^{-1}\\
-D^{-1}C(A-BD^{-1}C)^{-1}&D^{-1}+D^{-1}C(A-BD^{-1}C)^{-1}BD^{-1}
\end{array}
\right)\,,
\end{equation}
with
\begin{itemize}
\item $(A-BD^{-1}C)=
\left(
\begin{array}{cc}
\delta_a^{\phantom{a}b}&0\\
2\sum_I\bar{\theta}^{I\dot{a}}\theta_I^{b}&\delta^{\dot{a}}_{\phantom{\dot{a}}\dot{b}}
\end{array}
\right)$
\item $(A-BD^{-1}C)^{-1}=
\left(
\begin{array}{cc}
\delta_a^{\phantom{a}b}&0\\
-2\sum_I\bar{\theta}^{I\dot{a}}\theta_I^{b}&\delta^{\dot{a}}_{\phantom{\dot{a}}\dot{b}}
\end{array}
\right)$
\item $-(A-BD^{-1}C)^{-1}BD^{-1}=
-\left(
\begin{array}{c}
0\\
2i\bar{\theta}^{J\dot{a}}
\end{array}
\right)$
\item $-D^{-1}C(A-BD^{-1}C)^{-1}=
-(2i\theta_I^{b},0)$
\item $D^{-1}+D^{-1}C(A-BD^{-1}C)^{-1}BD^{-1}=
1_{\mathcal{N}\times\mathcal{N}}$
\end{itemize}
Hence:
\begin{equation}
\label{theta factor in the product, inverse}
\mathcal{U}^{-1}(0,\{\theta_{Ia}\},\{\bar{\theta}^{J\dot{a}}\})_A^{\phantom{A}B}=
\left(
\begin{array}{ccc}
\delta_a^{\phantom{a}b}&0&0\\
-2\sum_I\bar{\theta}^{I\dot{a}}\theta_I^{b}&\delta^{\dot{a}}_{\phantom{\dot{a}}\dot{b}}&-2i\bar{\theta}^{J\dot{a}}\\
-2i\theta_I^{b}&0&\delta_I^{\phantom{I}J}
\end{array}
\right)\,.
\end{equation}
The complete inverse transformation is then the product of 
(\ref{space-time factor in the product, inverse}) and (\ref{theta factor in the product, inverse}):
\begin{equation}
\label{explicit full transformation in 4d superspace, inverse}
\mathcal{U}^{-1}(x^\mu,\theta_{Ia},\bar{\theta}^{J\dot{a}})_A^{\phantom{A}B}=\left(
\begin{array}{ccc}
\delta_a^{\phantom{a}b}&0&0\\
-i\bar{y}^\mu(\bar{\sigma}_\mu)^{\dot{a}b}&\delta^{\dot{a}}_{\phantom{\dot{a}}\dot{b}}&-2i\bar{\theta}^{J\dot{a}}\\
-2i\theta_I^{b}&0&\delta_I^{\phantom{I}J}
\end{array}
\right)\,,
\end{equation}
where $\bar{y}^\mu=x^\mu-i\sum_I\theta_I\sigma^\mu\bar{\theta}^I=(y^\mu)^\dagger$ and we have again replaced, for each $I$, 
$\bar{\theta}^{I\dot{a}}\theta_I^{b}=\frac{1}{2}\theta_I\sigma^\mu\bar{\theta}^I(\bar{\sigma}_\mu)^{\dot{a}b}$. 
This is enough to derive the set of equations (\ref{barred 4d superspace}).

\section{Pascal Pyramid and \texorpdfstring{$\mathcal{N}=4$}{} chiral superfield}
\label{Appendix Pascal pyramid}

\subsection{Pascal Pyramid}
\begin{figure}[ht]
\begin{center}
\includegraphics[width=13cm]{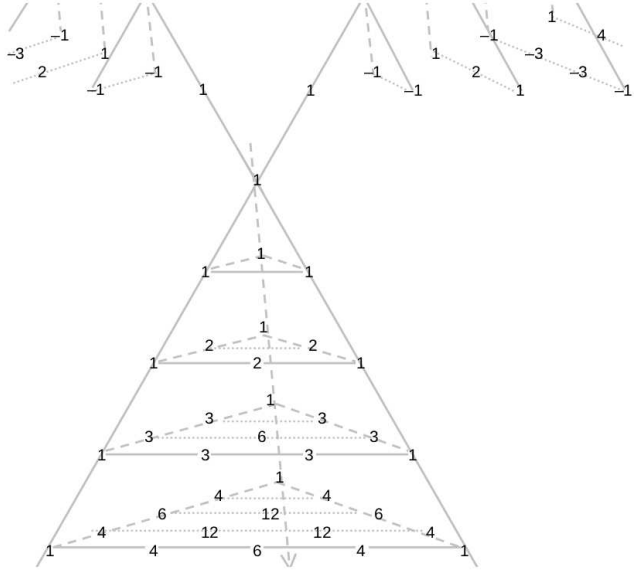}
\caption{\small A few layers in the positive Pascal pyramid. 
The pyramid contains in principle also negative extensions, but they have been cut here.
This picture has been taken from \cite{PyramidPicture}.}
\label{pyramid}
\end{center}
\end{figure}
In number theory, in analogy to the binomial expansion, whose coefficients can be organized into a triangle, 
one can consider the trinomial:
\begin{equation}
\label{trinomial expansion}
(a+b+c)^n=\sum_{m=0}^n\sum_{k=0}^m
\left(
\begin{array}{c}
n\\m 
\end{array}
\right) 
\left(
\begin{array}{c}
m\\k 
\end{array}
\right) 
a^{n-m}b^{m-k}c^{k}
\equiv\sum_{\substack{i,j,k\\i+j+k=n}}
\left(
\begin{array}{c}
n\\i,j,k 
\end{array}
\right)a^ib^jc^k\,,
\end{equation}
where the integer numbers appearing in the Pascal pyramid (tetrahedron) are just the coefficients 
$\left(
\begin{array}{c}
n\\i,j,k 
\end{array}
\right)=
\frac{n!}{i!j!k!}$
of the expansion (see figure \ref{pyramid}) of the trinomial. 
The power $n$ indicates the layer of the pyramid.

The pyramid has a huge number of symmetries and interesting properties.
For instance, by looking at specific examples or by using some combinatorics, one can check that there are 
$(n+1)(n+2)/2$ elements at each layer. 
Also, the sum of the entries of each layer is $3^n$, as can be seen by taking $a=b=c=1$ in (\ref{trinomial expansion}).
Moreover, by superimposing two consecutive layers, it is possible to prove that each entry at layer $n$ 
is given by the sum of all the entries at layer $n-1$ that surround it.
This is simply a property of the trinomial coefficients.

In number theory, one generalizes binomials and trinomials to multinomials and, in a similar way, we speak of an $m$-simplex,
which generalizes triangles and tetrahedrons. A multinomial is by definition 
\begin{equation}
(a_1+\dots+a_m)^n=
\sum_{\substack{k_1,\dots,k_m\\k_1+\dots+k_m=n}}
\left(
\begin{array}{c}
n\\k_1,\dots,k_m
\end{array}
\right)a_1^{k_1}\cdots a_m^{k_m}\,,
\end{equation}
where the coefficients are given by
\begin{equation}
\left(
\begin{array}{c}
n\\k_1,\dots,k_n 
\end{array}
\right)=
\frac{n!}{k_1!\dots k_n!}\,.
\end{equation}

\subsection{ \texorpdfstring{$\mathcal{N}=4$}{} chiral superfield }
In (\ref{pascal pyramid for N=2 chiral superfield}) we have written down the Pascal pyramid at layer two
and show that its elements are related to the number of certain kind of fields arising as components 
of an $\mathcal{N}=2$ chiral supermultiplet. 
Here we will do the same for $\mathcal{N}=4$ chiral supermultiplets and will find the pyramid at layer four.

As we have done in Section \ref{section: the chiral sector}, 
a six-dimensional $\mathcal{N}=4$ chiral superfield can be expanded in the theta variables as 
\begin{eqnarray}
\label{6d chiral N=4}
\Phi(X_{\alpha\beta},\theta_{I\alpha},\varphi_{IJ})&=&
A+\sum_I \theta_{I\alpha}\lambda^{I,\alpha}+\sum_{I<J}\theta_{I\alpha}\theta_{J\beta}E^{IJ,\alpha\beta}
+\sum_{I}\theta_{I\alpha}\theta_{I\beta}E^{I,\alpha\beta}
\nonumber\\
&&+\frac{1}{6}\sum_{I,J,K,L}\epsilon_{IJKL}\theta_{J\alpha}\theta_{K\beta}\theta_{L\gamma}\chi^{I,\alpha\beta\gamma}
+\sum_{I\neq J}\theta_{J\alpha}\theta_{J\beta}\theta_{I\gamma}\chi^{IJ,\alpha\beta\gamma}
\nonumber\\
&&+\theta_{1\alpha}\theta_{2\beta}\theta_{3\gamma}\theta_{4\delta}B^{\alpha\beta\gamma\delta}
+\frac{1}{2}\sum_{I,J,K,L}\epsilon_{IJKL}\theta_{I\alpha}\theta_{J\beta}\theta_{K\gamma}\theta_{L\delta}\tilde{B}^{IL,\alpha\beta\gamma\delta}
\nonumber\\
&&+\sum_{I<J}\theta_{I\alpha}\theta_{I\beta}\theta_{J\gamma}\theta_{J\delta}B^{IJ,\alpha\beta\gamma\delta}
+\frac{1}{6}\sum_{I,J,K,L}
\epsilon_{IJKL}\theta_{I\alpha}\theta_{I\beta}\theta_{J\gamma}\theta_{K\delta}\theta_{L\eta}\xi^{I,\alpha\beta\gamma\delta\eta}
\nonumber\\
&&+\frac{1}{2}\sum_{I\neq J\neq K}
\theta_{J\alpha}\theta_{J\beta}\theta_{K\gamma}\theta_{K\delta}\theta_{I\eta}\xi^{IJK,\alpha\beta\gamma\delta\eta}
\nonumber\\
&&+\frac{1}{2}\sum_{\substack{I<J\\I\neq J\neq K\neq L}}
\theta_{I\alpha}\theta_{J\beta}\theta_{K\gamma}\theta_{K\delta}\theta_{L\eta}\theta_{L\varepsilon}
C^{IJKL,\alpha\beta\gamma\delta\eta\varepsilon}
\nonumber\\
&&+\frac{1}{6}\sum_{I\neq J\neq K}
\theta_{I\alpha}\theta_{I\beta}\theta_{J\gamma}\theta_{J\delta}\theta_{K\eta}\theta_{K\varepsilon}
C^{IJK,\alpha\beta\gamma\delta\eta\varepsilon}
\nonumber\\
&&+\frac{1}{6}\sum_{I\neq J\neq K\neq L}
\theta_{J\alpha}\theta_{J\beta}\theta_{K\gamma}\theta_{K\delta}\theta_{L\eta}\theta_{L\varepsilon}\theta_{I\sigma}
\omega^{I,\alpha\beta\gamma\delta\eta\varepsilon\sigma}
\nonumber\\
&&+\theta_{1\alpha}\theta_{1\beta}\theta_{2\gamma}\theta_{2\delta}\theta_{3\eta}\theta_{3\varepsilon}\theta_{4\sigma}\theta_{4\rho}
D^{\alpha\beta\gamma\delta\eta\varepsilon\sigma\rho}+\dots\,.
\end{eqnarray}
Here, $I,J,\dots,{\rm etc.}=1,\dots,4$ can be raised/lowered with the flat (identity) metric, 
all the component fields are functions of $(X_{\alpha\beta},\varphi_{IJ})$ 
and the dots in the end represent quantities that will vanish on the light-cone. 
The four-dimensional superfield as well as its components are defined on the light-cone
and will be functions of $y^\mu=x^\mu+\sum_I\theta_I\sigma^\mu\bar{\theta}^I$ only. Moreover,
the $\varphi_{IJ}$-dependence is removed by replacing $\varphi_{IJ}=2iX^+\theta_I\cdot\theta_J$ 
and expanding everything in $\theta_{Ia}$.
The result is:
\begin{eqnarray}
\label{N=4 chiral multiplet expansion}
\Phi(y^\mu,\theta^{Ia})&\equiv&(X^+)^\Delta \Phi(X_{\alpha\beta},\theta_{I\alpha},\varphi_{IJ})=\nonumber\\
&&A(y)+\sum_I\theta_{Ia}\Lambda^{I,a}(y)+\sum_{I<J}\theta_{Ia}\theta_{Jb}F^{IJ,ab}(y)+\sum_I\theta_I^2 F^I(y)
\nonumber\\
&&+\frac{1}{6}\sum_{I,J,K,L}\epsilon_{IJKL}\theta_{Ja}\theta_{Kb}\theta_{Lc}\Upsilon^{I,abc}(y)
+\sum_{I\neq J}\theta_{Ia}\theta_{J}^2 \Upsilon^{IJ,a}(y)
+\theta_{1a}\theta_{2b}\theta_{3c}\theta_{4d}B^{abcd}
\nonumber\\
&&+\frac{1}{2}\sum_{\substack{J<K\\J\neq K\neq L}} \theta_{Ja}\theta_{Kb}\theta_L^2 B^{JKL,ab}(y)
+\sum_{I<J}\theta_I^2\theta_J^2 B^{(IJ)}(y)
\nonumber\\
&&+\frac{1}{6}\sum_{I,J,K,L}\epsilon_{IJKL}\theta_I^2\theta_{Ja}\theta_{Kb}\theta_{Lc}\Xi^{I,abc}(y)
+\frac{1}{2}\sum_{I\neq J\neq K}\theta_{Ia}\theta_J^2\theta_K^2\Xi^{I(JK),a}(y)
\nonumber\\
&&+\frac{1}{2}\sum_{\substack{I<J\\I\neq J\neq K\neq L}}\theta_{Ia}\theta_{Jb}\theta_K^2\theta_L^2 C^{IJ(KL),ab}(y)
+\frac{1}{6}\sum_{I\neq J\neq K}\theta_I^2\theta_J^2\theta_K^2 C^{(IJK)}(y)
\nonumber\\
&&+\frac{1}{6}\sum_{I\neq J\neq K\neq L}\theta_{Ia}\theta_J^2\theta_K^2\theta_L^2 \Omega^{I,a}(y)
+\theta_1^2\theta_2^2\theta_3^2\theta_4^2 D(y)\,.
\end{eqnarray}
The fractional coefficients are such that each field appears only once in the sum.
The four-dimensional components are expressed in terms of the six-dimensional ones appearing in (\ref{6d chiral N=4}).
For example, for the first few components we have: 
\begin{eqnarray}
A(y)&=& (X^+)^\Delta A(X,0)\nonumber\\
\Lambda^{Ia}(y)&=& (X^+)^{\Delta+1}\Big[\lambda^{Ia}(X,0)-iy_\mu(\bar{\sigma}^\mu)^{\dot{a}a}\lambda^I_{\dot{a}}(X,0)\Big]\nonumber\\
F^{IJ,ab}(y)&=& (X^+)^{\Delta+1}\Big[iX_{\alpha\beta}\epsilon^{ab}E^{IJ,\alpha\beta}_{\rm aver.}(X,0)
-4i\epsilon^{ab}\frac{\partial A_{\rm aver.}}{\partial \varphi_{IJ}}(X,0)\Big]\nonumber\\
F^{I}(y)&=& (X^+)^{\Delta+1}\Big[-iX_{\alpha\beta}E^{I,\alpha\beta}(X,0)+2i\frac{\partial A}{\partial\varphi_{II}}(X,0)\Big]\nonumber\\
\Upsilon^{I,abc}(y)&=& (X^+)^{\Delta+2}\Big[\epsilon^{ab}iX_{\alpha\beta}\chi_{\rm aver.}^{I,\alpha\beta c}(X,0)
+\epsilon^{ab}X^{\alpha\beta}y_\mu(\bar{\sigma}^\mu)^{\dot{a}c}\epsilon_{\dot{a}\dot{c}}\chi_{\rm aver.}^{I,\alpha\beta\dot{c}}(X,0)\nonumber\\
&&\phantom{(X^+)^{\Delta+2}}
-24i\epsilon^{bc}\epsilon_{IJKL}\frac{\partial \lambda^{Ja}}{\partial\varphi_{KL}}(X,0)
-24\epsilon_{IJKL}\epsilon^{bc}y_\mu(\bar{\sigma}^\mu)^{\dot{a}a}\frac{\partial\lambda^J_{\dot{a}}}{\partial\varphi_{KL}}(X,0)\Big]\nonumber\\
\Upsilon^{I,a}(y)&=& (X^+)^{\Delta+2}\Big[-iX_{\alpha\beta}\chi^{IJ,\alpha\beta a}(X,0)
-X_{\alpha\beta}y_\mu(\bar{\sigma}^\mu)^{\dot{c}a}\epsilon_{\dot{c}\dot{d}}\chi^{IJ,\dot{d}}(X,0)
+2i\frac{\partial\lambda^{Ia}}{\partial\varphi_{IJ}}(X,0)\nonumber\\
&&\phantom{(X^+)^{\Delta+2}}
+y_\mu(\bar{\sigma}^\mu)^{\dot{a}a}\frac{\partial\lambda^I_{\dot{a}}}{\partial\varphi_{JJ}}(X,0)
-i\frac{\partial\lambda^{Ja}}{\partial\varphi_{IJ}}(X,0)
-y_\mu(\bar{\sigma}^\mu)^{\dot{a}a}\frac{\partial\lambda^J_{\dot{a}}}{\partial\varphi_{IJ}}(X,0)\Big]\nonumber\\
\dots&&\dots\qquad\qquad\dots\qquad\dots\qquad\dots
\end{eqnarray}
etc. Here the averaged quantities are defined whenever $I\neq J$ as done in the main text, e.g.
\begin{equation}
\theta_{I\alpha}\theta_{J\beta}E^{IJ,\alpha\beta}(X,\varphi)=
-iX_{\alpha\beta}\theta_I\cdot\theta_J E^{IJ,\alpha\beta}_{\rm aver.}(X,\varphi)\,,
\end{equation}
and should be regarded as shorter labels of longer expressions.

Let us now count the number of component fields contained in the multiplet (\ref{N=4 chiral multiplet expansion}). There are:
\begin{itemize}
\item one scalar field $A(y)$
\item four spinors $\Lambda^I(y)$ in the representation (1/2,0)
\item $3\times4/2=6$ rank-2 tensors $F^{IJ}(y)$, $I<J$ in the representation (1,0)
\item four scalars $F^I(y)$
\item four spinors $\Upsilon^I(y)$ in the representation (3/2,0)
\item $4\times3=12$ spinors $\Upsilon^{IJ}(y)$, $I\neq J$, in the representation (1/2,0)
\item one rank-4 tensor $B(y)$ in the representation (2,0)
\item $4\times3=12$ rank-2 tensors $B^{JKL}(y)$, $J<K$ and $J\neq K\neq L$, in the representation (1,0)
\item $4\times3/2=6$ scalars $B^{(IJ)}(y)$, $I<J$ and symmetric in $I,J$
\item four spinors $\Xi^I(y)$ in the representation (3/2,0)
\item $4\times(3\times2/2)=12$ spinors $\Xi^{I(JK)}$, $J\neq K\neq L$ and symmetric in $K,L$, in the representation (1/2,0)
\item six rank-2 tensors $C^{IJ(KL)}(y)$, $I<J$ and symmetric in $K,L$, in the representation (1,0)
\item $4\times3\times2/3!=4$ scalars $C^{(IJK)}(y)$, symmetric in $I,J,K$
\item four spinors $\Omega^I(y)$, in the representation (1/2,0)
\item one scalar $D(y)$.
\end{itemize}
This field content can be summarized by using the Pascal pyramid at layer $\mathcal{N}=4$:
\begin{equation}
\begin{array}{ccccccccc}
1&&4&&6&&4&&1\\
&4&&12&&12&&4\\
&&6&&12&&6&&\\
&&&4&&4&&&\\
&&&&1&&&&
\end{array}
\Longleftrightarrow
\begin{array}{ccccccccc}
A&&\Lambda^I&&F^{IJ}&&F^I&&B\\
&\Upsilon^I&&\Upsilon^{IJ}&&B^{JKL}&&\Xi^I\\
&&B^{(IJ)}&&\Xi^{I(JK)}&&C^{IJ(KL)}&&\\
&&&C^{(IJK)}&&\Omega^{I}&&&\\
&&&&D&&&&
\end{array}
\end{equation}
The total number of fields is given by the sum of the elements of the pyramid and amounts, as it should be, to $3^4=81$.

\end{document}